\documentclass[3p,preprint,twocolumn]{elsarticle}

\usepackage{lineno,hyperref}
\usepackage{bm}
\usepackage{placeins}
\usepackage{array}
\usepackage{caption}
\usepackage{upgreek}

\usepackage{booktabs,makecell,multirow,array,colortbl,dcolumn,stfloats} 

\usepackage{framed} 
\usepackage{multicol} 
\usepackage{nomencl} 
\makenomenclature
\renewcommand*\nompreamble{\begin{multicols}{2}}
\renewcommand*\nompostamble{\end{multicols}}

\modulolinenumbers[5]

\usepackage{color}
\usepackage{relsize}
\usepackage[utf8]{inputenc}
\usepackage{mathtools}
\usepackage{amsmath}
\usepackage{amssymb}
\usepackage[size=small]{todonotes} 
\usepackage{tcolorbox} 
\usepackage{pdflscape} 
\usepackage{geometry} 
\usepackage{setspace}\usepackage{threeparttable} 
\usepackage{pdfpages}
\usepackage{lipsum}
\newcommand{\ve}[1]{\boldsymbol{#1}} 
\newcommand{\te}[1]{\boldsymbol #1} 

\newcommand{\M}[1]{\underline{\underline{\boldsymbol{#1}}}} 
\newcommand{\V}[1]{\underline{\boldsymbol{#1}}} 
\newcommand{\diffp}[2]{\frac{\partial #1}{\partial #2}} 
\newcommand{\inte}[3]{\mathop{\int}_{ #1} #2 \; \mathrm{d} #3} 
\newcommand{\intee}[4]{\mathop{\int}_{ #1}^{ #2} #3 \; \mathrm{d} #4} 

\newcommand{\tr}{\mathrm{tr}\,}
\newcommand{\dev}{\mathrm{dev}}

\newcommand{\calB}{{\cal B}}

\newcommand{\calH}{{\cal H}}
\newcommand{\calF}{{\cal F}}

\newcommand{\Gc}{G_\mathrm{c}}
\newcommand{\sig}{\te{\sigma}}

\newcommand{\eps}{\te{\varepsilon}}
\newcommand{\epse}{\te{\varepsilon}^\mathrm{e}}
\newcommand{\epsp}{\te{\varepsilon}^\mathrm{p}}

\newcommand{\psie}{\psi^\mathrm{e}}
\newcommand{\psip}{\psi^\mathrm{p}}

\newcommand{\Def}{\coloneqq }

\journal{Theoretical and Applied Fracture Mechanics}

\usepackage{hyperref}
\usepackage{xurl}
\hypersetup{breaklinks=true}

\begin{document}

\begin{frontmatter}

\title{Phase-field models for ductile fatigue fracture}


\author[mymainaddress]{Martha Kalina}

\author[mymainaddress]{Tom Schneider}

\author[mysecondaddress]{Haim Waisman}

\author[mymainaddress,mythirdaddress]{Markus K\"{a}stner\corref{mycorrespondingauthor}}
\cortext[mycorrespondingauthor]{Corresponding author}
\ead{markus.kaestner@tu-dresden.de}

\address[mymainaddress]{Chair of Computational and Experimental Solid Mechanics, Dresden University of Technology, Dresden, Germany}
\address[mysecondaddress]{Department of Civil Engineering and Engineering Mechanics, Columbia University, New York City, USA}
\address[mythirdaddress]{Dresden Center for Computational Materials Science (DCMS), Dresden University of Technology, Dresden, Germany}

\begin{abstract} 
Fatigue fracture is one of the main causes of failure in structures. However, the simulation of fatigue crack growth is computationally demanding due to the large number of load cycles involved. Metals in the low cycle fatigue range often show significant plastic zones at the crack tip, calling for elastic-plastic material models, which increase the computation time even further. In pursuit of a more efficient model, we propose a simplified phase-field model for ductile fatigue fracture, which indirectly accounts for plasticity within the fatigue damage accumulation. Additionally, a cycle-skipping approach is inherent to the concept, reducing computation time by up to several orders of magnitude. 

Essentially, the proposed model is a simplification of a phase-field model with elastic-plastic material behavior. As a reference, we therefore implement a conventional elastic-plastic phase-field fatigue model with nonlinear hardening and a fatigue variable based on the strain energy density, and compare the simplified model to it. Its approximation of the stress-strain behavior, the neglect of the  plastic crack driving force and consequential range of applicability are discussed. 

Since in fact the novel efficient model is similar in its structure to a phase-field fatigue model we published in the past, we include this older version in the comparison, too. Compared to this model variant, the novel model improves the approximation of the plastic strains and corresponding stresses and refines the damage computation based on the Local Strain Approach. 
For all model variants, experimentally determined values for elastic, plastic, fracture and fatigue properties of AA2024 T351 aluminum sheet material are employed.
\end{abstract}


\end{frontmatter}

\section{Introduction} 

\begin{figure}
	\def\svgwidth{\linewidth}\small{
		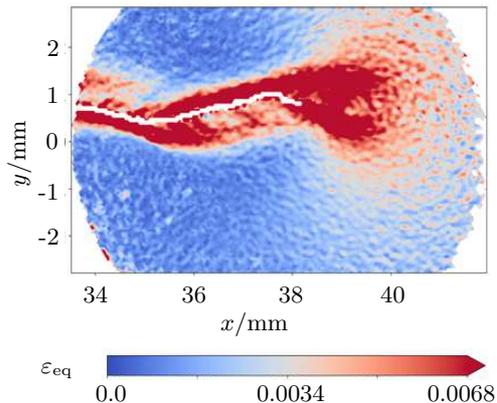}
	\caption{Measured strain field at the tip of a fatigue crack. $\varepsilon_\mathrm{eq}$ is the von Mises equivalent stress. Taken from \cite{paysan_integration_2022} recorded with Microscopic Digital Image Correlation (MDIC) \cite{paysan_robotassisted_2023}.
	}\label{fig:DIC}
\end{figure}

Fatigue fracture is one of the most frequent reasons why engineering structures fail \cite{stephens_metal_2000}, yet modeling fatigue crack \textit{growth} still poses a major challenge in computational mechanics. This is due to both the complexity of the fatigue phenomenon on the microscale as well as the high number of load cycles to be simulated, which commonly leads to high computational effort. Especially in metals, fatigue is caused by plasticity on different scales \cite{bathias_fatigue_2010}. In low cycle fatigue (LCF), higher load amplitudes lead to significant plastic effects in regions prone to stress concentrations like notches and the crack tip. But even in high cycle fatigue (HCF), when the component remains in the elastic range on a macroscopic level, plastic micro deformations occur on the grain scale and cause damage that ultimately lead to failure. Crack initiation is caused by plastic slip bands, which occur at dislocations, and merge into micro and later macro cracks. Macroscopic fatigue crack growth is controlled by the complex interplay of the initial plastic zone, re-plastification and crack closure. Fig.~\ref{fig:DIC} shows a digital image correlation (DIC) image of a tip of a fatigue crack in aluminum sheet material, whose distribution of equivalent strain hints at the plastic zone. Also, plasticity causes effects in the crack growth rate during complex loading paths, also known as \textit{effects of sequence}. E.g. exceptionally high loads within the loading sequence can lead to widening of the plastic zone, reducing the crack propagation rate in the following load cycles \cite{bathias_fatigue_2010}.

This calls for an efficient simulation strategy for ductile fatigue crack growth. The phase-field method is eminently suited for this task due to its ability to model crack initiation, propagation and sudden residual fracture in a straightforward way. Arbitrary crack paths in complex components can be simulated without auxiliary conditions, which is a decisive advantage over other crack modeling strategies, such as element deletion, XFEM or cohesive zone models. Based on the pioneering works of Francfort, Marigo and Bourdin \cite{francfort_revisiting_1998,bourdin_numerical_2000,bourdin_variational_2008}, who cast the fracture problem as a variational framework and introduced the regularized description of the crack, Miehe et al. \cite{miehe_thermodynamically_2010,miehe_phase_2010} introduced the phase-field model for brittle fracture with a widely used implementation strategy.

The generality of the model allowed for numerous extensions of the phase-field framework: See  \cite{dammass_unified_2021} for a review on phase-field models for \textit{viscous} behavior and \cite{alessi_comparison_2018} on models considering \textit{ductile} fracture. While most ductile phase-field models such as \cite{miehe_phase_2016} employ a degraded plastic energy density, forcing the term to be part of the crack driving force, there are other approaches of coupling the phase-field to plastic quantities, e.\,g. via the degradation function in \cite{ambati_phasefield_2015}. Models especially focused on plastic shear bands are \cite{mcauliffe_coupled_2016} and \cite{mcauliffe_unified_2015}.

Just as well, numerous phase-field models have been published for \textit{fatigue} fracture under cyclic loading. To this end, see our overview \cite{kalina_overview_2023} of models for fatigue fracture including both brittle and ductile behavior. In that paper, we classified these models in mainly two groups according to their integration of fatigue in the previously static model: Depending on an accumulating fatigue variable, either the fracture toughness of the material is gradually reduced or the crack driving force increased in order to allow for sub-critical crack growth. In terms of the fatigue variable itself, mainly two variants are used: First, the strain energy density (SED) is accumulated to account for the stress history the material has undergone during cyclic loading. Often, these models employ cycle jump techniques \cite{cojocaru_simple_2006,heinzmann_adaptive_2024}, which extrapolate the fatigue variable over several load cycles in order to save computation time. Another group of models uses empiric concepts from the field of structural durability, which impose certain assumptions on the damaging effects of load cycles based on cyclic experiments. Extrapolation of the fatigue variable is part of these concepts and can be applied within the modeling framework, given the crack propagation rate is small enough for the stress-strain state to change slowly.

Some of the phase-field fatigue models published to date are also suitable for ductile fracture behavior. Most of them introduce a plastic energy density term, e.\,g. due to hardening, which contributes to the crack driving force \cite{seles_numerical_2020,ulloa_phasefield_2021,khalil_phasefield_2021,haveroth_nonisothermal_2020}. Given relatively high loading amplitudes (LCF), which cause macroscopic plastic strains, and a suitable elastic-plastic material model, fatigue crack growth can be modeled without an additional fatigue variable, just from the accumulating plastic SED alone, which gradually increases the crack driving force. However, usually, these elastic-plastic phase-field models are supplemented by a fatigue variable based on the elastic SED \cite{carrara_framework_2019,aldakheel_phasefield_2022,hasan_phasefield_2021} in order to also cover fatigue fracture in HCF where the whole component behaves macroscopically elastic. Some models even include the plastic SED in the fatigue variable, next to the crack driving force, doubling the effect of plasticity on crack growth \cite{ulloa_phasefield_2021,khalil_phasefield_2021}.

Even with the help of cycle jump strategies, the computational cost for phase-field models with elastic-plastic material behavior is high: Still, a significant number of load cycles has to be resolved during the simulation. For complex structures, this is hardly feasible in practically relevant situations. Therefore, simplified models for ductile fatigue fracture can be a reasonable alternative for fully elastic-plastic material models.

Previously, we introduced such a reduced model with a fatigue variable based on a heuristic approach from structural durability \cite{seiler_efficient_2020}, which reduces the material's fracture toughness under repetitive loading. The model includes assumptions for the stress-strain behavior based on cyclic stress-strain curves recorded for the material, which are valid as long as plasticity occurs only locally to a limited extent. The material's resistance against fatigue damaging, on the other hand, is characterized by single-amplitude cyclic tests, relating load amplitude and fatigue life. The model was mainly motivated by  the interpretation of the fatigue variable as a damage variable in the sense of classic structural durability concepts.

In this paper, we take a different turn to derive an efficient, pseudo-plastic model. We start from a phase-field model with conventional elastic-plastic material model with nonlinear hardening. The fatigue variable is based on the elastic and plastic SED -- one of the most common and generic choices. We then simplify this model by approximating the stress-strain behavior according to Glinka's revaluation strategy \cite{glinka_energy_1985}, requiring cyclic stress-strain data of the material as an input. The fatigue damage accumulation for each load cycle by integrating the SED is simplified by the Heitmann damage parameter \cite{heitmann_life_1984}, which requires only minimal stress and strain data for each load cycle. In this way, we avoid the explicit simulation of load cycles, benefiting from the type of cycle jump scheme inherent to the approach. As a result, the proposed model overcomes the extensive elastic-plastic material model coupled to the  phase-field fatigue model, alleviating the significant computation time. For clarity, we term the fully elastic-plastic reference model \textcircled{\smaller A}, the novel efficient model \textcircled{\smaller B} and our former model \cite{seiler_efficient_2019} model \textcircled{\smaller C}.

We show that the novel model \textcircled{\smaller B} is indeed an enhanced version of \textcircled{\smaller C}, as it improves the approximation of the elastic-plastic stress-strain behavior and the assessment of the damaging effect of stress-strain hysteresis through the new damage parameter. We compare the novel \textcircled{\smaller B} model not only with its predecessor, but also with the fully coupled elastic-plastic phase-field model \textcircled{\smaller A}. 
The latter is used as a reference to evaluate the quality of  model \textcircled{\smaller B} by studying the approximated stresses and neglecting the plastic contributions in crack driving force and fatigue variable.
For all model types, fracture toughness, elastic and plastic parameters as well as  characteristic cyclic material curves are taken from experiments with AA 2024 T351 aluminum sheet material.  

The paper is structured as follows. At first, in Section \ref{sec:framework}, we set up a general modeling framework suited to describe all model types. Within this framework, in Section \ref{sec:elastplast}, the  elastic-plastic phase-field fatigue model \textcircled{\smaller A} is introduced. The simplified pseudo-plastic phase-field model \textcircled{\smaller B} for fatigue fracture is presented in Section \ref{sec:LSA}, which we compare to the former model \textcircled{\smaller C}. Finally, Section \ref{sec:numex} compares all model variants and discusses their applicability using numerical examples.

\section{General framework for phase-field fatigue models including plasticity}
\label{sec:framework}

To begin with, we set up a general framework for phase-field models for fatigue fracture, in which all models presented in the following operate in. The integration of  plasticity in the phase-field model follows \cite{seles_numerical_2020,seles_general_2021,seles_microcrack_2021}, while the derivation via the principle of virtual power follows the work by \cite{khalil_generalised_2022}.

We consider a domain $\calB \subset \mathbb{R}^n$ and its boundary $\partial\calB$. The material points are described by their location $\ve{x}$ and their displacement $\ve{u}(\ve{x},t)$ at time $t$. Due to the typically low fatigue loads, the total strain $\eps(\ve{x},t)$ is assumed to be small and can be divided additively into elastic and plastic strain  contributions $\epse(\ve{x},t)$ and $\epsp(\ve{x},t)$, respectively:
\begin{equation}
	\eps \Def  \frac{1}{2} \left( \nabla\ve{u} + \nabla\ve{u}^\top \right) = \epse + \epsp.
\end{equation} 
The phase-field variable $d(\ve{x},t)$ describes the crack topology in a regularized manner, whereby $d=1$ marks a fully developed crack and $d=0$ intact material. Cyclic stressing of the material leads to fatigue damage, which is described by the scalar fatigue variable $\calF(\ve{x},t)$. 
Considering all these variables, the total energy density is given as
\begin{multline} \label{eq:SED}
	W \Def W_{\mathrm{el}}(\epse,d) + W_{\mathrm{pl}}(\eps,d,\epsp,\dot{\eps}^\mathrm{p})\\\ + W_\mathrm{frac,fat}(d,\nabla d;\calF) .
\end{multline} 
It consists of an elastic part $W_\mathrm{el}$, a plastic part $W_\mathrm{pl}$ and a combined fracture and fatigue term $W_\mathrm{frac,fat}$, which are defined as
\begin{align}
	 W_{\mathrm{el}} & \Def g(d)\, \psie_+(\epse) + \psie_-(\epse)  \\
	 W_{\mathrm{pl}} & \Def g(d)\, \psip(\eps,\epsp,\dot{\eps}^\mathrm{p})   \label{eq:Wpl}\\
	 W_\mathrm{frac,fat} & \Def h(\calF)\, \Gc \gamma(d,\nabla d). \label{eq:fracfat}
\end{align}
Thereby, the SEDs are defined as
\begin{align}
	\psie_+ & \Def  \frac{1}{2} K \langle \tr\epse \rangle_+^2 + \mu \, \tr\left(\dev\epse\cdot\dev\epse \right)  \\
	\psie_- & \Def \frac{1}{2} K \langle \tr\epse \rangle_-^2 \\
	\psip & \Def \intee{0}{t}{\left(\dev(\sig^*)-\te{\upchi}\right):\dot{\eps}^\mathrm{p}}{\tau}
\end{align}
with constant $K=\lambda+\frac{2}{3}\mu$ depending on the Lamé parameters $\lambda$ and $\mu$ and $\langle\cdot\rangle_\pm$ being the Macaulay bracket.
The elastic SED $W_\mathrm{el}$ is split into tensile and compressive parts. Here we use the volumetric-deviatoric split by Amor et al. \cite{amor_regularized_2009}. Both $W_\mathrm{el}$ and $W_\mathrm{pl}$ are degraded by the phase-field via the degradation function $g(d) = (1-d)^2$. $\sig^*$ is defined as the undegraded stress, $\te{\upchi}$ is the backstress tensor in case of kinematic hardening. The fracture energy is defined by the material parameter fracture toughness $\Gc$ and the regularized crack surface density 
\begin{equation}\label{eq:gamma}
	\gamma_\ell=\left( \frac{d^2}{2\ell} + \frac{\ell}{2}\nabla d\cdot\nabla d \right)
\end{equation}
with characteristic length $\ell$.
Under cyclic loading, the fracture toughness is degraded gradually by the fatigue degradation function $h(\calF)$. Given the SED (\ref{eq:SED}), 
the Cauchy stress and the conjugate forces for the phase-field $\zeta_d,\zeta_{\nabla d}$ are
\begin{equation} \label{eq:conjug}
	 \diffp{W}{\epse} =:  \sig \quad  -\diffp{W}{d} =: \zeta^d \quad -\diffp{W}{\nabla d} =: \zeta^{\nabla d}.
\end{equation}
Accordingly, the stress can be split additively into
\begin{equation}
	\sig(\epse) \Def \diffp{W_\mathrm{el}}{\epse} = g(d) \sig_+(\epse) + \sig_-(\epse),
\end{equation}
whereby the  undamaged stress is 
\begin{equation}
	\sig^*(\epse) \Def \sig_+(\epse) + \sig_-(\epse).
\end{equation}

With the virtual velocities $\delta\dot{\ve{u}}, \delta\dot{d}$, the principle of virtual power can be set up as
\begin{multline} \label{eq:PovP}
	\inte{\calB}{\left[\sig:\nabla\delta\dot{\ve{u}} + \zeta^d\,\delta\dot{d} + \zeta^{\nabla d}\cdot\nabla\delta\dot{d}\right]}{v} = \\\ 
	\inte{\mathcal{B}}{\ve{f}\cdot\delta\dot{\ve{u}}}{v} + \inte{\partial\mathcal{B}^\mathrm{N}}{\ve{t}\cdot\delta\dot{\ve{u}}}{a},
\end{multline}
with $\ve{f}$  being volume force and and $\ve{t}$ the traction vector on the Neumann boundary $\partial\calB^\mathrm{N}$, respectively. Applying the Gauss divergence theorem, equation (\ref{eq:PovP}) under consideration of (\ref{eq:gamma}) yields the balance of momentum, the evolution equation for the phase-field, as well as boundary conditions for the coupled problem, leading to 
\begin{equation}
	\begin{array}{r}
		\nabla\cdot\sig + {\ve{f}}  = \ve{0} \\
		\nabla\cdot\zeta^{\nabla d} - \zeta^d  = 0
	\end{array}
	 \text{ in } \calB, \quad
	 \begin{array}{r}
	 	\sig\cdot\ve{n}  = {\ve{t}}\text{ on } \partial\calB^\mathrm{N} \\
	 	\nabla d\cdot\ve{n}  = 0 \text{ on } \partial\calB.
	 \end{array}
\end{equation}
With the definitions of $\zeta^d,\zeta{\nabla d}$ in (\ref{eq:conjug}), the phase-field evolution equation becomes
\begin{multline} 
	0 = \Gc{h(\calF)}\left(\ell\Delta d - \frac{d}{\ell}\right) { + \Gc\ell\,\nabla d\,\nabla h(\calF)} \\\ -g'(d)  \left(\psie_+ + \psip \right).
\end{multline}
We adopt the heuristic approach to ensure irreversibility of fracture by Miehe et al. \cite{miehe_phase_2010}
\begin{multline} \label{eq:evo}
	0 = \Gc{h(\calF)}\left(\ell\Delta d - \frac{d}{\ell}\right) { + \Gc\ell\nabla d\nabla h(\calF)} \\\ -g'(d)\underbrace{  \max_{\tau\in[0,t]} \left(\psie_+(\tau) + \psip(\tau) \right)}_{\calH}
\end{multline}
by introducing a history variable $\calH$.
See \cite{gerasimov_penalization_2019} for an alternative approach based on a penalty parameter.

\section{Elastic-plastic phase-field model for fatigue fracture \textcircled{\smaller A}}
\label{sec:elastplast}

The straightforward approach to include ductile material behavior in a phase-field fracture model is to employ a full elastic-plastic material model. Here, we adopt an Armstrong-Frederick \cite{armstrong_mathematical_1966,chaboche_modelization_1979} type of plasticity model with nonlinear kinematic and isotropic hardening. This model is known to be suitable for reversed and cyclic loading \cite{lemaitre_mechanics_1998}. The model is parametrized for an AA2024 T351 aluminum sheet material.
Later, this model will serve as a reference for the pseudo-plastic model \textcircled{\smaller B} presented in Section \ref{sec:LSA}. 

\subsection{Theory}

\subsubsection*{Elastic-plastic material model}

We adopt a von Mises yield criterion with the yield function
\begin{equation}
	f^\mathrm{p} \Def \sqrt{\frac{3}{2}||\dev(\sig)-{\te{\upchi}}||^2} - \sigma^\mathrm{y} 
\end{equation}
with the yield stress $\sigma^\mathrm{y}$ and the backstress tensor $\te{\upchi}$.
The flow rule according to associated plasticity is
\begin{equation}
	\dot{\eps}^\mathrm{p} = \dot{\varepsilon}^\mathrm{p}_\mathrm{eq} \diffp{f^\mathrm{p}}{\sig}
\end{equation}
with equivalent plastic strain rate and plastic strain
\begin{equation}
	\dot{\varepsilon}^\mathrm{p}_\mathrm{eq} = \sqrt{\frac{2}{3}\dot{\eps}^\mathrm{p}:\dot{\eps}^\mathrm{p} }, \quad \varepsilon^\mathrm{p}_\mathrm{eq} = \intee{0}{t}{\sqrt{\frac{2}{3}\dot{\eps}^\mathrm{p}:\dot{\eps}^\mathrm{p}}}{\tau}.
\end{equation}
Isotropic hardening with a Voce hardening law is described by the following yield stress
\begin{equation}
	\sigma^\mathrm{y} = \sigma_0 + \bar Q \left(1-e^{-\bar{b}\,\varepsilon^p_\mathrm{eq} }\right) 
\end{equation}
with $\sigma_0$ being the initial yield stress and $\bar Q,\, \bar b$ hardening parameters.
The three backstress tensors $\te{\upchi}_k$ of kinematic hardening evolve according to
\begin{multline}
	\te{\upchi} = \sum_k 	\te{\upchi}_k, \quad \dot{\te{\upchi}}_k = \frac{2}{3} C_k \dot{\varepsilon}^\mathrm{p}_\mathrm{eq} \diffp{f^\mathrm{p}}{\sig} - \gamma_k \dot{\varepsilon}^\mathrm{p}_\mathrm{eq} \te{\upchi}_k \\\ =  \frac{2}{3} C_k \dot{\eps}^\mathrm{p}  - \gamma_k \dot{\varepsilon}^\mathrm{p}_\mathrm{eq} \te{\upchi}_k.
\end{multline}
Three backstress tensors is generally seen as the minimum number, yet it already reproduced the nonlinear behavior well, as the parametrization will show. The set of parameters $C_k,\gamma_k$ for each backstress tensor characterizes its contribution to kinematic hardening. This law for the evolution for backstress and plastic strain defines the plastic energy term $\psip$ in (\ref{eq:Wpl}). In this way, hardening terms enter the crack driving force in the phase-field evolution equation (\ref{eq:evo}) and thereby contribute to promote crack growth in case of plastic yielding.


\subsubsection*{Fatigue variable and fatigue function}

In order to enable fatigue crack growth for  subcritical cyclic loads, eq. \ref{eq:SED} introduced a fatigue variable $\calF$. A fatigue degradation function $h(\calF)$ degrades the fracture toughness $\Gc$ gradually according to (\ref{eq:fracfat}). Here, we adopt the fatigue variable being the accumulated SED
\begin{equation}
	\calF(t) = \frac{1}{\vartheta_N}\intee{0}{t}{\mathrm{H}(\vartheta\dot{\vartheta})\vartheta\dot{\vartheta}}{\tau}
\end{equation}
with the scaling parameter $\vartheta_N$ as in \cite{carrara_framework_2019}. It accumulates both elastic and plastic SED through the variable
\begin{equation}
	 \vartheta(\tau)=g(d)\,\left(\psie_+(\epse(\tau)) +\psip({\epse(\tau)},\epsp(\tau))\right)
\end{equation}
as in \cite{ulloa_phasefield_2021}.
The Heaviside step function $\mathrm{H}(\cdot)$ ensures accumulation under loading only, excluding unloading. 
During LCF, which usually leads to localized macroscopic plastic effects, the crack is dominated by $\psip$ through the crack driving force and the fatigue variable. In contrast, in the HCF regime, in the absence of plasticity, the crack is driven by $\psie_+$, solely through the fatigue variable. This fatigue variable can account for mean stress effects \cite{carrara_framework_2019}: Load cycles of the same stress or strain amplitude have a more significant fatigue contribution if they are shifted towards the tensile range, thereby their mean stress becoming larger.

The fatigue degradation function is set as
\begin{equation} \label{eq:fatdeg}
	h(\calF)  = (1-h_0)(1-\calF)^\xi + h_0
\end{equation}
with the parameters $h_0$ and $\xi$.

\subsection{Parametrization}

This section describes the determination of the parameters of the elastic plastic material model as well as the fracture toughness with standardized experiments. The parameters of the fatigue degradation function are not subject to standardized experiments and need to be calibrated subsequently.

\subsubsection*{Elastic-plastic material model}

\begin{figure} 
	\def\svgwidth{\linewidth}\small{
		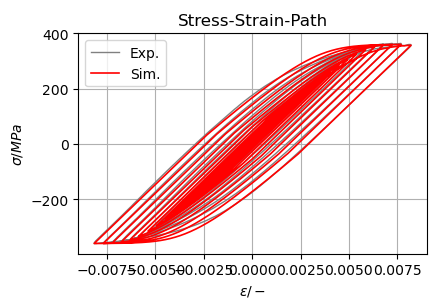}
	\caption{Model \textcircled{\smaller A}: Calibration of a Armstrong-Frederick-type plasticity model with 3 backstress tensors to low-cycle experimental data of aluminium AA2024 T351 sheet material.}
	\label{fig:Chaboche_fit}
\end{figure}

The elastic-plastic material model is parametrized using strain-controlled uniaxial tension-compression experiments. Dogbone specimen were cut from the AA2024 T351 aluminum sheet material. A Teflon-coated buckling brace was installed. The specimen was loaded with 17 load cycles of decreasing amplitudes.  The acquired stress-strain data is displayed in Fig. \ref{fig:Chaboche_fit}. It also shows the result of the parameter calibration using a Basin-hopping optimization \cite{wales_global_1997} for inverse parametrization. The determined plastic parameters, including those for nonlinear isotropic hardening $\bar Q, \bar b$ and nonlinear kinematic hardening, $C_k, \gamma_k$ for each of the three backstress tensors $\te{\upchi}_k$, are listed in Tab. \ref{tab:pars}. 

\subsubsection*{Fracture toughness}

The fracture toughness of the aluminum material was determined through Compact Tension (CT) tests in \cite{kalina_fatigue_2023}, where the experimental procedure is described in detail. Therein, the material was tested  in several sets of experiments with different orientations, as the material shows a slight anisotropy in fracture resistance. Here, we neglect these deviations and apply the constant value $\Gc=0.0153$ kN\,mm$^{-1}$.

\begin{figure*}
	\def\svgwidth{\linewidth}\small{
		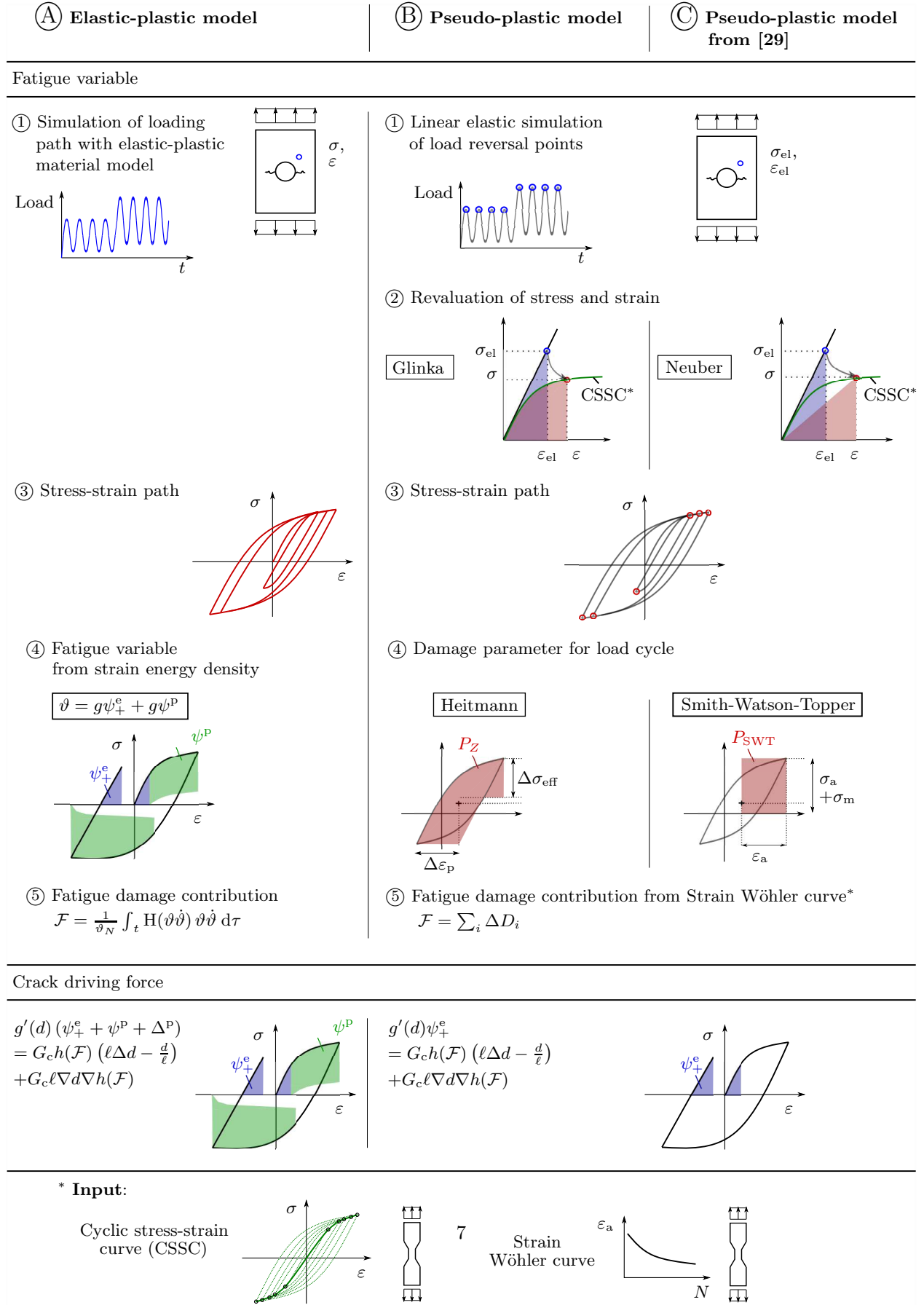}
	\caption{Overview of the model variants compared in this paper: The elastic-plastic model \textcircled{\smaller A} with its conventional elastic-plastic material model serves a reference for the two simplified, pseudo-plastic models \textcircled{\smaller B} and \textcircled{\smaller C}. They simplify the determination of the stress-strain path and the computation of the fatigue variable, incorporating material data from cyclic experiments. Model \textcircled{\smaller B} is the main subject of this paper, model \textcircled{\smaller C} has been published in \cite{seiler_efficient_2020}.
	}\label{fig:models}
\end{figure*}

\section{New pseudo-plastic phase-field model for fatigue fracture \textcircled{\smaller B}}
\label{sec:LSA}

The huge computational effort of elastic-plastic models -- including model \textcircled{\smaller A} described in the previous section -- is mainly due to the return mapping algorithm, which is carried out at each integration point. This calls for simplified models and acceleration techniques. In this section, we introduce a phase-field model for fatigue loading with approximated elastic-plastic behavior  as a simplification of model  \textcircled{\smaller A} presented in Section \ref{sec:elastplast}. This new model is termed model \textcircled{\smaller B}. Both \textcircled{\smaller A} and \textcircled{\smaller B} are visualized in Fig.~\ref{fig:models}.

\subsection{Key idea}

Starting from the ductile phase-field model \textcircled{\smaller A} with nonlinear hardening, we introduce three measures in order to simplify the model and reduce computation time:
\begin{itemize}
	\item An approximation of the stress-strain behavior based on cyclic stress-strain data of the material replaces the elastic-plastic material model with hardening laws. The effectiveness of the approximation is studied in Section \ref{sec:stressplastzone}
	\item With the stress-strain path approximated by a closed form analytical expression, the fatigue parameter $\calF$ can be directly computed from the known hysteresis curve. At the same time, the fatigue evaluation is enriched with experimental data of the material.
	\item The plastic contribution $\psip$ to the crack driving force $\calH$ in (\ref{eq:evo}) is neglected. Section \ref{sec:driving} shows that this term is indeed negligible for many loading scenarios. 
\end{itemize}
In this context, the total energy density simplifies to
\begin{equation} \label{eq:SED_LSA}
	W = W_{\mathrm{el}}(\epse,d) +  W_\mathrm{frac,fat}(d,\nabla d;\calF) .
\end{equation} 
The elastic-plastic material model with  its energy contribution $W_\mathrm{pl}=g(d)\psip$ is removed completely, avoiding the costly return-mapping iteration. The resulting crack driving force is compared in Fig. \ref{fig:models} for models \textcircled{\smaller A} and \textcircled{\smaller B}. Naturally, this procedure is only valid for  plastic processes which remain local and take place in a small plastic zone around the crack tip. Section \ref{sec:numex} studies the applicability of the concept. 

Since only the load reversal points are necessary for the computation of $\calF$, this simulation strategy is even more efficient than an elastic-plastic model \textcircled{\smaller A} with a cycle jump approach. In contrast to the latter, not a single complete load cycle has to be simulated, as will be explained later.

The novel approach \textcircled{\smaller B} is in fact an extension of our phase-field fatigue model \cite{seiler_efficient_2019} based on the Local Strain Approach (LSA) \textcircled{\smaller C} \cite{seeger_grundlagen_1996}. Now, the stress-strain approximation according to Neuber \cite{neuber_theory_1961} is replaced by the more enhanced approach by Glinka \cite{glinka_energy_1985}\footnote{ Tang et al. \cite{tang_classical_2024} incorporated a Glinka rule in their phase-field model, but haven't compared it to an elastic-plastic model yet}. Further, instead of the damage parameter by Smith, Watson and Topper \cite{smith_stressstrain_1970} $P_\mathrm{SWT}$, the parameter by Heitmann et al.
\cite{heitmann_life_1984} $P_Z$ is used, which approximates the SED more accurately. Model \textcircled{\smaller C} is also explained in Fig. \ref{fig:models}. 

Still, with these changes taken, the novel model operates within the framework of the LSA. The modularity of this approach allows for a interchangeability of different model aspects between models \textcircled{\smaller B} and \textcircled{\smaller C}. 
In the following, the novel parts of the model, i.\,e. the approximation of the fatigue behavior and the fatigue variable, are explained in detail.

\subsection{Approximation of stress-strain behavior}

Assuming plastic zones at notches and around the crack tip are small and the entire component behaves mostly elastic, the cyclic material behavior can be approximated according to Glinka \cite{glinka_energy_1985}. They presume that the SED  remains the same compared to the solely elastic solution, once small scale yielding occurs at a notch, dominated and controlled by elastic surroundings. Therefore, in a first step, stresses and strains are computed assuming elastic material behavior, see Fig. \ref{fig:models} \textcircled{\smaller 1}. Then, the stress-strain path is revaluated, assuming it follows the cyclic stress-strain curve (CSSC). The latter is determined experimentally and explained in Section \ref{sec:LSAparam}. The CSSC is described by the Ramberg-Osgood ansatz \cite{ramberg_description_1943} 
\begin{equation} \label{eq:CSSC}
	\varepsilon_\mathrm{a} = \frac{\sigma_\mathrm{a}}{E} + \left( \frac{\sigma_\mathrm{a}}{K'} \right)^{1/n'}
\end{equation}
and is determined from the reversal points of the stress-strain path in this cyclic test.

Following Glinka, the SED for the elastic solution, depending on $\varepsilon_\mathrm{el},\sigma_\mathrm{el}$, is set equal to the SED for the pseudo-plastic behavior with $\varepsilon,\sigma$. For the latter, the SED is integrated from the CSSC. See Fig. \ref{fig:models} \textcircled{\smaller 2} for a visualization of the stress-strain revaluation: The blue (elastic SED) and red area (pseudo-plastic SED) are set to be equal, resulting in
\begin{equation}
	\frac{\sigma_\mathrm{el}^2}{2E} = \frac{\sigma^2}{2E} + \frac{\sigma}{n'+1}\left(\frac{\sigma}{K'}\right)^{1/n'}
\end{equation}
given the pseudo-plastic material behavior from (\ref{eq:CSSC}). In order to recover the stress-strain path for the whole loading sequence \textcircled{\smaller 3}, assumptions for the behavior during reloading have to be made, compensating for the lack of flow rules and hardening laws as in an elastic-plastic material model: According to Masing \cite{masing_eigenspannungen_1926}, the stress and strain range during reloading is twice the amplitude value of the first loading. Complementary rules for material memory for variable amplitude loading, including rules for re- and subsequent loading were set up in \cite{clormann_rainflowhcm_1985}. As a result, the stress-strain path of the material point is available for the  entire loading history.

Since the CSSC is determined from uniaxial experiments, we employ the von Mises equivalent stresses $\sigma_{\mathrm{el}}$ and $\sigma$ throughout the stress revaluation within the 3D material routine. In the crack tip area, it is dominated by the stress component in crack opening direction.
We investigate the quality of the stress revaluation in this context in Section~\ref{sec:stressplastzone}.

\subsection{Fatigue variable and damage parameter}

With the approximated stress-strain path as an input, the fatigue variable is then determined for each hysteresis with the help of a damage parameter, see Fig. \ref{fig:models} \textcircled{\smaller 4}. Here, we employ the damage parameter by Heitmann $P_Z$ \cite{heitmann_life_1984}. Analogously to the fatigue variable of model \textcircled{\smaller A}, which consists of the elastic SED and the dissipated plastic SED integrated as part of the area within the hysteresis, $P_Z$ consists of an effective elastic $W_\mathrm{e,eff}$ and a plastic part $W_\mathrm{p}$
\begin{equation}
	P_Z = f_1 W_\mathrm{e,eff}+f_2 W_\mathrm{p} = 2.9\, W_\mathrm{e,eff}+2.5\, W_\mathrm{p}.
\end{equation}
This makes it ideal for the application in the context of crack tip plasticity. Multiplied by the crack length $a$ it is equal to the cyclic $J$-integral \cite{tanaka_cyclicjintegral_1983}. A numerical solution of the latter yields the coefficients $f_1$ and $f_2$ \cite{heitmann_life_1984}. As they follow from the $J$-integral, they are general and therefore independent of the material. See Fig. \ref{fig:models} \textcircled{\smaller 4} for a graphic comparison of the damage variable from \textcircled{\smaller B} and the fatigue variable in  \textcircled{\smaller A}.
Since the shape of the hysteresis is known through the CSSC as a function (\ref{eq:CSSC}), the plastic part can be specified as
\begin{equation}
	 W_\mathrm{p} = \frac{1}{1+n'}\Delta\sigma\Delta\varepsilon^\mathrm{p} = \frac{1}{1+n'}\,\Delta\sigma \left(\frac{\Delta\sigma}{2K'}\right)^{1/n'}.
\end{equation}
For the elastic part
\begin{equation}
	W_\mathrm{e,eff} = \frac{\Delta\sigma_\mathrm{eff}^2}{2E},
\end{equation} 
only the stress and strain range beyond the point of crack closure is considered, excluding the phase when the crack faces are in contact.
Experiments show that the proportion of crack closure compared to the total stress range depends on the crack length, but converges towards the value for long cracks \cite{radaj_ermuedungsfestigkeit_2007}. 
Hence, Heitmann applies this value in the form of an effective stress range  due to closure according to Schijve \cite{schijve_formulas_1981}
\begin{equation}
	\Delta\sigma_\mathrm{eff} = \,3.72(3-R)^{-1.74}\,\Delta\sigma.
\end{equation}
This function of the stress ratio $R$ is a general relation that was derived independently of the material. It was validated experimentally, also for aluminum AA2024 \cite{schijve_formulas_1981} specifically, among others. The integration of the $R$-ratio allows to consider the influence of the mean stress as well. This is especially important since the evaluation of the damaging effect later on is conducted with fatigue data (strain Wöhler curves) for the material. With the help of the damage parameter, the fatigue data recorded for a specific experimental setup can be employed for other stress ratios as well, with the damage parameter scaling the fatigue damage influence of the load accordingly.

These strain Wöhler curves are applied to characterize the material's resistance against fatigue damage. They are recorded with a set of constant-amplitude strain-controlled fatigue tests, as described in Section \ref{sec:LSAparam}. The original relation between applied strain amplitude $\varepsilon_\mathrm{a}$ and number of load cycles until crack initiation $N_\mathrm{f}$
\begin{equation} \label{eq:SWC}
	\varepsilon_\mathrm{a} = \frac{\sigma'_\mathrm{f}}{E} \,(2N_\mathrm{f})^b + \varepsilon'_\mathrm{f}\,(2N_\mathrm{f})^c
\end{equation}
is reformulated \cite{nihei_evaluation_1986}, replacing $\varepsilon_\mathrm{a}$ with the damage parameter $P_Z$
\begin{equation} \label{eq:dwl_pz}
	P_Z=\sigma_\mathrm{f}'(2N_\mathrm{f})^b \left[ 0.64467\frac{\sigma_\mathrm{f}'}{E}(2N_\mathrm{f})^b+\frac{10}{1+n'}\varepsilon_\mathrm{f}'(2N_\mathrm{f})^c \right].
\end{equation}
Once $P_Z$ is known for a load cycle $i$, (\ref{eq:dwl_pz}) yields $N_{\mathrm{f}i}$. The fatigue damage contribution for this load cycle can then be computed from 
\begin{equation}
	\Delta D_i=1/N_{\mathrm{f}i}.
\end{equation}
Finally, the current fatigue variable is the sum of fatigue damage contributions the material point has seen so far
\begin{equation}
	\calF=\sum_i \Delta D_i.
\end{equation}
The fatigue variable degrades the fracture toughness of the material through the fatigue degradation function $h(\calF)$ (\ref{eq:fatdeg}), just like in model \textcircled{\smaller A}.

Compared to the formerly in model \textcircled{\smaller C} used damage parameter $P_\mathrm{SWT}$ by Smith, Watson and Topper \cite{smith_stressstrain_1970}, the Heitmann parameter $P_Z$ approximates the dissipated energy more accurately (cf. Fig. \ref{fig:models} \textcircled{\smaller 4}), considers crack closure effects  and is also shown to predict more accurate fatigue lives for components \cite{heitmann_life_1984}. Due to its consideration of both the elastic SED and the plastic strain range, it is suitable for both LCF and HCF. By the involvement of material fatigue life curves, the concept can include valuable material information into the fatigue damage computation. However, only material strain Wöhler curves are employed, no crack growth data is necessary.

\subsection{Implications for computational time}

Our proposed methodology computes the fatigue damage from the stress amplitude or stress range exclusively and therefore no simulation of the loading and unloading phase of the load cycle is necessary. Instead, only reversal points are simulated to record the stress state. Also, the load reversal points are the critical states of the load cycle for the crack to grow.
If the crack growth rate is small, the fatigue damage contribution for several load cycles can be extrapolated, mimicking a simplified cycle jump procedure. This reduces the number of simulation increments even further.

\subsection{Parametrization}
\label{sec:LSAparam}

The pseudo-plastic "material model" in models \textcircled{\smaller B} and \textcircled{\smaller C} can be directly parametrized through cyclic experiments. This is not the case for the parameters of the fatigue degradation function, which are still subject to calibration.
are parametrized with experimental data for the same AA2024 T351 aluminum sheet material as the conventional model \textcircled{\smaller A}. Apart from the fracture toughness, which was already determined in the previous section, two sets of cyclic experiments have to be conducted:
The LSA is based on two material characteristics that describe the cyclic stress-strain behavior an the fatigue resistance of the material, respectively: The CSSC and the strain Wöhler curve. Both are determined with uniaxial tension-compression tests according to the norm SEP 1240 \cite{_testing_2006}. Dogbone shaped specimens cut in 0°, 45° and 90° angle to the rolling direction are used. The resulting curves  show almost no dependence on the direction, though. The chosen stress ratio between maximum and minimum load $R=F_{\min}/F_{\max}=-1$ requires a buckling brace since the load is in both tensile and compressive range.

\subsubsection*{Stress-strain behavior}
The CSSC is determined from incremental step tests (IST) \cite{landgraf_determination_1969}. Strain-controlled load cycles are applied  with block-wise increasing and decreasing amplitudes. The maximum strain amplitude $\varepsilon_\mathrm{a}$ for the 0° and 90° direction is 0.8 \% and in 45° 0.65 \%. The reversal points for each loading block characterize the stress-strain behavior at the respective point in the lifetime of the specimen. Here, we adhere to the standard implementation of the Local Strain approach, employing a single CSSC after half the lifetime of the material. Nonetheless, if the material undergoes a substantial change in behavior throughout the lifespan of the specimens or if it exhibits notable deviations between tensile and compressive range, we can incorporate the modification proposed in \cite{kuhne_fatigue_2018}. This modification accommodates tension-compression asymmetry and accounts for the transient evolution of the stress-strain relationship. The parameters of the Ramberg-Osgood formulation \cite{ramberg_description_1943} of the CSSC (\ref{eq:CSSC}) are $K'=750$ MPa and $n'=0.0728$. The underlying experimental results and the calibration are shown in Fig. \ref{fig:WoehlerCSSC}a.

\subsubsection*{Fatigue resistance}
\begin{figure} 
	\def\svgwidth{\linewidth}\small{
		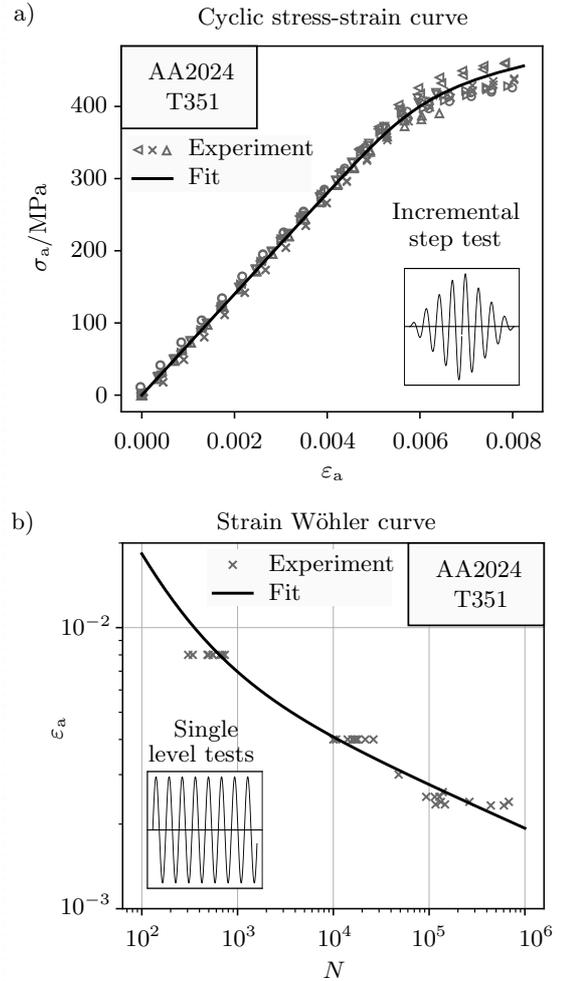}
	\caption{\textbf{(a)} Cyclic stress-strain curve (CSSC) from incremental step test (IST) and \textbf{(b)} S-N  (Wöhler) curve recorded with single level tests. Tensile test specimen from aluminum sheet material AA2024 T351 were used. Data was acquired for different orientations of the sheet material and, in case of the IST, for load sequences starting with both tensile and compressive loading. Data first published in \cite{kalina_fatigue_2023}. Both curves serve as an input for models \textcircled{\smaller B} and \textcircled{\smaller C}.}
	\label{fig:WoehlerCSSC}
\end{figure}

Similar to the Wöhler or $S$-$N$ curve, the strain Wöhler curve characterizes the material's resistance against cyclic loading, with the main difference that here the cyclic tests are strain-controlled. Constant amplitude tests with different strain amplitudes are used to determine the relation between strain amplitude $\varepsilon_\mathrm{a}$ and the number of endured load cycles $N_\mathrm{f}$. The calibration of the ansatz (\ref{eq:SWC}) yielded the parameters \mbox{$\sigma'_\mathrm{f}=1206$\,MPa}, \mbox{$\varepsilon'_\mathrm{f}=0.9975$}, \mbox{$b=-0.152$} and \mbox{$c=-0.860$}. Experimental results and calibration are displayed in Fig. \ref{fig:WoehlerCSSC}b.

\section{Validation example for novel model \textcircled{\smaller B}}
\label{sec:validation}

The novel pseudo-plastic model \textcircled{\smaller B} is now validated with a numerical example, showing its ability to model fatigue crack growth. To this end, we simulate the crack growth experiment that led to the MDIC image in Fig.~\ref{fig:DIC}.

\subsection{Simulation setup}

All model variants are implemented in the FE software ABAQUS using the UMAT interface. Model \textcircled{\smaller A} is solved with a "staggered" approach according to \cite{navidtehrani_simple_2021}. Their static elastic phase-field implementation served as a template here. Models \textcircled{\smaller B} and \textcircled{\smaller C} were solved with a "monolithic" approach, as \cite{navidtehrani_simple_2021} term it, which practically complies with the popular \textit{Alternate Minimization} scheme.\footnote{In the literature, the terms \textit{monolithic} and \textit{staggered} are used for a variety of different solution schemes. Here, we stick to the labeling by \cite{navidtehrani_simple_2021}, since we also built upon their implementation. The following table specifies the residual vectors $\V{r}^u,\,\V{r}^d$ for displacement and phase-field and the tangent with submatrices $\M{K}^{uu},\,\M{K}^{dd}$:
	\vspace{0.5cm}
	\begin{tabular}{lll}
		&  "Monolithic"    & "Staggered"    \\  
		Residual  & 	
		$\V{r}^u(^t_i\V{u},^t_{i-1}\V{d})$ & 
		$\V{r}^u(^{t}_i\V{u},^t_{i-1}\V{d})$\\ 
		& 
		$\V{r}^d(^t_{i-1}\V{u},^t_i\V{d})$ &
		$\V{r}^d(^{t-1}\V{u},^t_i\V{d})$   \\   
		Tangent & 
		$\left[\begin{array}{ll} \M{K}^{uu} & \M{0} \\ \M{0} & \M{K}^{dd} \end{array} \right]$ &  $\left[\begin{array}{ll} \M{K}^{uu} & \M{0} \\ \M{0} & \M{K}^{dd} \end{array} \right]$  \\ 
		& &
	\end{tabular}
}

\begin{figure}
	\def\svgwidth{\linewidth}\small{
		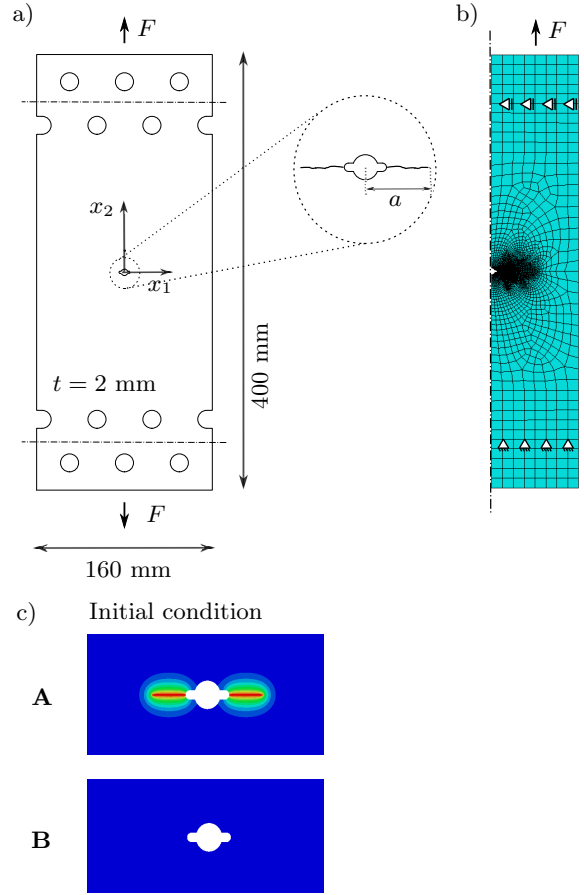}
	\caption{\textbf{(a)} Middle Tension (MT) specimen according to norm ASTM E674-15 \cite{astme647-15_standard_2015}. The crack length is indicated by $a$. \textbf{(b)} Mesh and boundary conditions. \textbf{(c)} Two setups are used in the simulations: With (\textbf{A}) and without (\textbf{B}) an initial crack indicated by fixed nodal phase-field values.
	}\label{fig:mtsetup}
\end{figure}

Simulations are conducted with a middle tension (MT) geometry displayed in Fig.~\ref{fig:mtsetup}. The MT160 specimen has width of 160\,mm and a thickness of 2\,mm.  The top and bottom clamping area are modeled as a rigid block. Due to symmetry, only half the geometry is modeled, with the intersection made along the $x_2$-axis. The geometry is meshed with quadratic elements with a local refinement. In the area of the potential crack, their dimension in planar direction is 0.4~mm with 5 elements in thickness direction. 
All material parameters are summarized in Tab.~\ref{tab:pars} and are applied in all following simulations, if not stated otherwise.

Two setup variants are displayed in  Fig.~\ref{fig:mtsetup}: Variant \textbf{A} specifies a pre-existing crack of $a=8$\,mm as an initial phase-field value $d=1$. Variant \textbf{B} has no initial crack, enabling the simulation to cover crack initiation, too. Both will be used in the subsequent simulations in this paper. In this chapter, however, only variant \textbf{B} without an initial crack is employed.

\begin{table*}
	\caption{Model parameters for all three model variants  for aluminum sheet material AA2024 T351.}\label{tab:pars}
	\renewcommand{\arraystretch}{2} 
	\begin{tabular}{l|c|c|c|c}
		\hline
		& \multicolumn{2}{c|}{\textbf{Elastic-plastic model \textcircled{\smaller A}}} & \multicolumn{2}{c}{\textbf{Pseudo-plastic models \textcircled{\smaller B}\textcircled{\smaller C}}}  \\ 
		\hline\noalign{\smallskip}
		\makecell{\textbf{Elastic} \\parameters} & \multicolumn{4}{c}{$E=67630$ MPa, $\nu=0.33$ } \\ \hline\noalign{\smallskip}
		\makecell{(Pseudo) \\ \textbf{Plastic} \\ parameters} & \multicolumn{2}{c|}{\makecell{Initial yield stress $\sigma^0=196.144$ MPa \\ Isotropic hardening: $\bar Q=0.279$ MPa, $\bar b=3.29$ \\ Kinematic hardening: $C_1=98340$ MPa, \\ $\gamma_1=1728.058$,  $C_2=64300$ MPa, $\gamma_2=2267.112$,\\$C_3=8.9280$ MPa, $\gamma_3=1127.010$}} & \multicolumn{2}{c}{\makecell{Cyclic stress-strain curve:\\ $K'=750$ MPa, $n'=0.0728$}}   \\ \hline\noalign{\smallskip}
		\makecell{\textbf{Fracture} \\parameters} & \multicolumn{4}{c}{ \makecell{$\Gc=0.0153$ kN\,mm$^{-1}$, {$\ell=2$ mm}}} \\ \hline\noalign{\smallskip}
		\makecell{\textbf{Fatigue} \\variable} & \multicolumn{2}{c|}{\makecell{$\vartheta_N=1$ MPa$^2$}} &  \multicolumn{2}{c}{\makecell{Strain Wöhler curve: \\ $\sigma'_f = 1206$ MPa, $\varepsilon'_f=0.9975$,\\ $b=-0.152$, $c=-0.860$}} \\
		\hline\noalign{\smallskip}
		\makecell{\textbf{Fatigue} \\ degradation \\ {function}} & \multicolumn{2}{c|}{\makecell{ $h_{\min}= 0.05$, $\xi=1$}} & \multicolumn{2}{c}{\makecell{ $h_{\min}= 0.001$, $\xi=1450$}}   \\ \hline
	\end{tabular}
\end{table*}

%
%

\subsection{Crack growth simulation}

\begin{figure}
	\def\svgwidth{\linewidth}\small{
		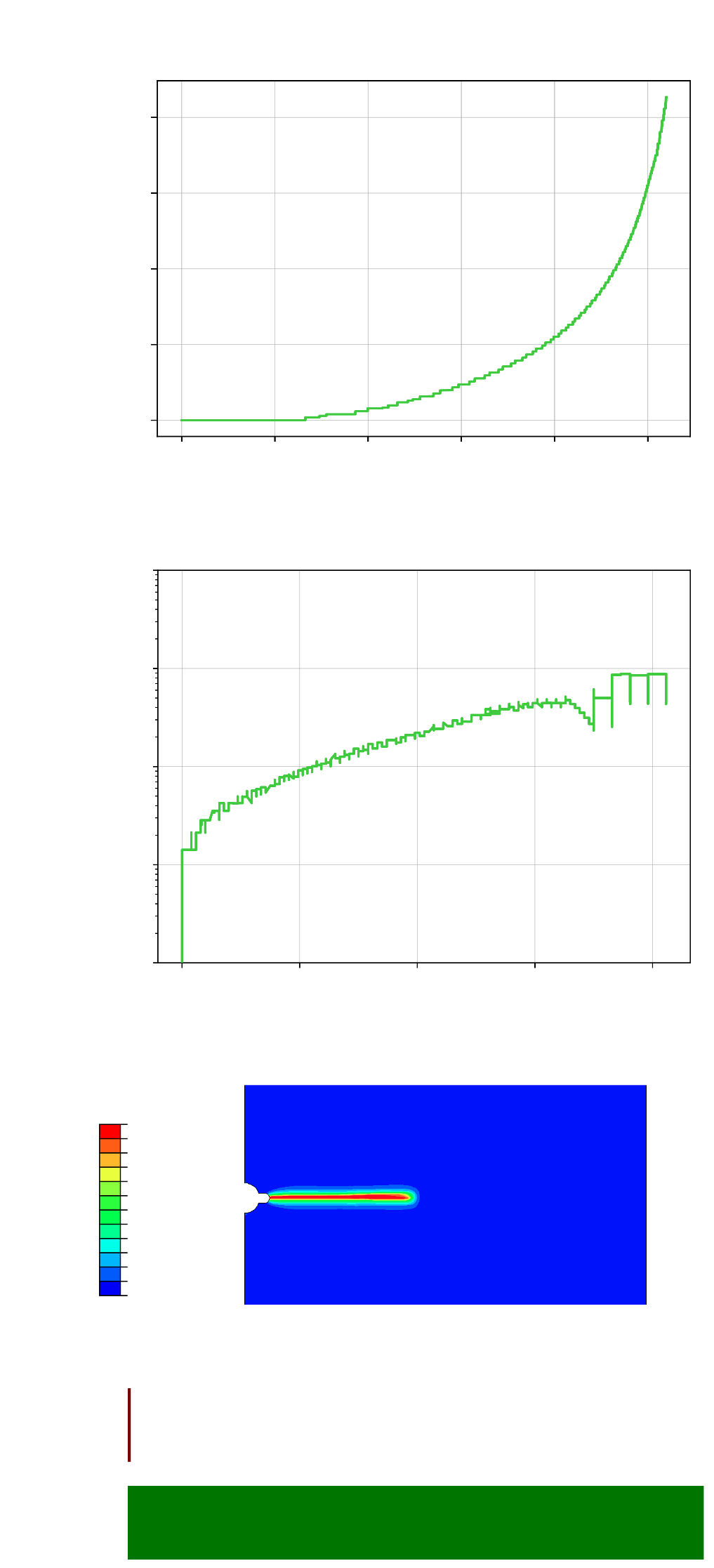}
	\caption{Simulation of fatigue crack growth with model \textcircled{\smaller B}. Parameters	$h_0=0.001,\xi=1450,\ell=0.5$\,mm are employed. \textbf{(a)} Crack length $a$ over load cycle number $N$ shows crack initiation phase and later crack growth with increasing speed. \textbf{(b)} Crack growth rate $\mathrm{d}a/\mathrm{d}N$ shows Paris-like stable crack growth phase. \textbf{(c)} Phase-field $d$ indicates crack. One half of the symmetric geometry is displayed. \textbf{(d)} Comparison of computational time for the first 2400 load cycles for models \textcircled{\smaller A} and \textcircled{\smaller B}.
	}\label{fig:sim_exp}
\end{figure}
The specimen is loaded with pseudo load cycles with a maximum force of $F_{\max}=15$\,kN and a load ratio of $R=0.1$. Parameters $h_0=0.001,\xi=1450$ of the fatigue degradation function and characteristic length $\ell=0.5$\, mm are adopted from earlier 2D simulations in \cite{kalina_fatigue_2023}. Starting with 2250 load cycles, the amount of load cycles per increment is adaptively reduced to 40 gradually throughout the simulation with a heuristic approach. 

The crack growth curve in Fig.~\ref{fig:sim_exp}a) demonstrates the model's ability to simulate crack initiation, which sets in after ca. 600\,000 load cycles. Subsequently, the graph shows the  typical acceleration of the crack growth. This is underlined by the Paris plot Fig.~\ref{fig:sim_exp}b), which features the typical phase of stable crack growth. Fig.~\ref{fig:sim_exp}c) displays the phase-field profile of the crack.

\subsection{Speed-up}



Due to the large number of load cycles, a simulation of the entire crack path or even significant crack growth was not feasible with the elastic-plastic model  \textcircled{\smaller A} for comparison. However, a simulation of the first 2400 load cycles yielded a \textbf{189 times longer simulation time} on average per load cycle. In this case, 10 increments per load cycle were used to resolve the loading path. While this is just a particular example, it underlines the tremendous potential in speed-up possible with efficient model \textcircled{\smaller B}.

One way to speed up simulations with traditional model \textcircled{\smaller A} would be cycle jump approaches \cite{cojocaru_simple_2006}. However, this is beyond the scope of this paper. In any case, a significant number of cycles has to be resolved during the simulation, which is still a disadvantage compared to the pseudo-plastic model  \textcircled{\smaller B} presented here. 
For a discussion of acceleration techniques for phase-field models in general, see \cite{kalina_overview_2023}. \cite{heinzmann_adaptive_2024} recently proposed a particularly effective cycle jump scheme.

The major advantage of the pseudo-plastic models  \textcircled{\smaller B} and  \textcircled{\smaller C} is that the number of increments required does not depend on the number of load cycles, but on the development of the fatigue variable and the crack itself.
For the same geometry and fracture pattern, but different load levels and therefore fatigue lives, the simulation time can be expected to be in the same order of magnitude.

\section{Numerical studies}
\label{sec:numex}

The novel pseudo-plastic model \textcircled{\smaller B} promises great advantages in computational time, as shown above. Its range of applicability is studied next.
To this end, its simplifications in terms of plasticity are validated by comparing it to the elastic-plastic model \textcircled{\smaller A}. Specifically, we investigate the quality of the stress revaluation and assess the omission of the plastic contribution to the crack driving force for distinct loading scenarios. The capability of the model to reproduce effects of the loading sequence for more complex loading paths is compared to the conventional elastic-plastic model, too. 

The following studies employ the same geometry and set-ups as the previous section.

\subsection{Stress and plastic zone}
\label{sec:stressplastzone}

A good reproduction of the stress distribution and the plastic zone is a key demand to the simplified model. Before presenting the numerical examples, we briefly discuss the literature on the quality of stress revaluation techniques in non-phase-field contexts. 

\subsubsection*{General discussion and literature}

In the novel model \textcircled{\smaller B}, the Glinka rule replaces the revaluation according to Neuber \cite{neuber_theory_1961} compared to model \textcircled{\smaller C}. Both schemes operate within the LSA framework, though, see also Fig. \ref{fig:models} \textcircled{\smaller 2}. In contrast to Neuber, Glinka does not neglect the plastic dissipation when setting up the energy equivalence. According to \cite{glinka_energy_1985}, Glinka reproduces experimentally determined inelastic strains at the notch quite accurately, whereas Neuber overestimates stresses and strains.
	
Both revaluation techniques have been compared to experimentally determined strains and FE simulations in various contexts, e.\,g. in \cite{knop_glinka_2000}. Therein, the authors confirm most previous findings for monotonic and cyclic loading: Under mostly plane strain conditions, Glinka reproduces experimental results more accurately than Neuber. In any loading case, Glinka estimates stresses the better, the more pronounced the stress concentration is -- which is particularly relevant for crack tips. In fact,  the difference between the two revaluation techniques becomes more pronounced for higher stress concentrations. The interaction with the regularized crack tip in phase-field fracture simulations now poses a new situation that is investigated in the following. 

\subsubsection*{Stress under monotonic loading}

To evaluate the quality of the stress revaluation according to Glinka and Neuber in models  \textcircled{\smaller B} and  \textcircled{\smaller C}, respectively, this section contrasts simulations of these models with the  elastic-plastic model  \textcircled{\smaller A}, which serves as a reference. A monotonic displacement up to $\bar{u}=0.8$\,mm is applied to the above-mentioned MT160 geometry. The monotonic load allows to study higher stress levels compared to a cyclic load. In order to compare the undegraded stress without the interference of a developing phase-field crack, crack growth is prevented by setting the fracture toughness to $10^4$ times the actual value of this material.

\begin{figure*}
	\def\svgwidth{\linewidth}\small{
		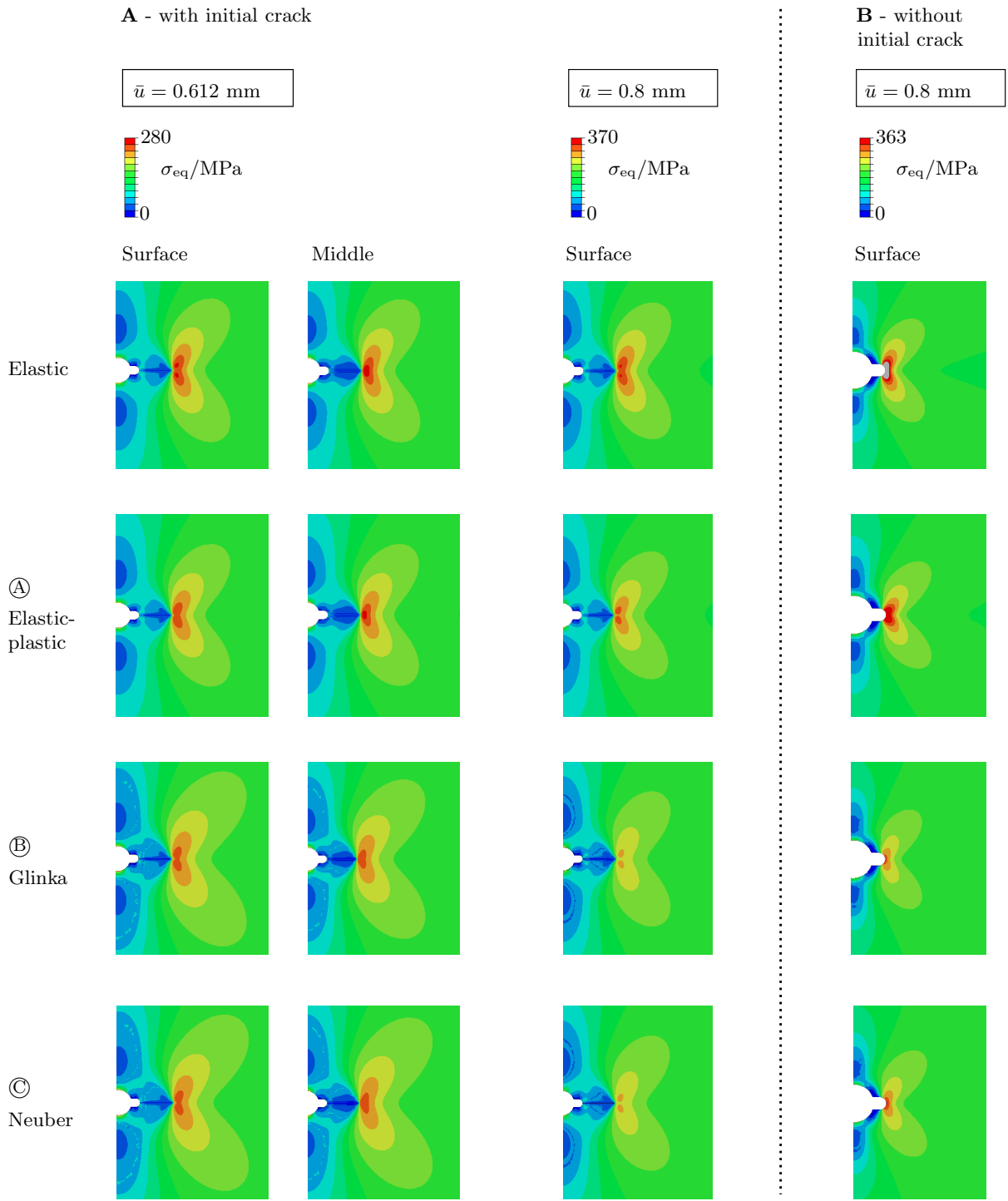}
	\caption{ Comparison of stress states. Two MT geometries (A - with initial phase-field crack and B - without it) are subject to \underline{monotonic} displacement load $\bar{u}$. Equivalent stress is displayed for the outer surface of the sheet material and for middle axis in case of specimen A. Compared are a purely elastic model, an elastic-plastic material model and the revaluation rules according to Glinka and Neuber. The fracture toughness is set a to a high value to prevent further degradation. 
	Stress redistribution takes place for the elastic-plastic model, hence, stresses are smaller compared to the elastic model (without stress revaluation). The approximations according to Glinka and Neuber yield qualitatively similar stress distributions, yet underestimate the maximum stress slightly.
	}\label{fig:stress}
\end{figure*}

Fig.~\ref{fig:stress} displays the stresses for two setups: with (\textbf{A}) and without (\textbf{B}) initial phase-field crack.  The Mises equivalent stress determined with  model \textcircled{\smaller A} serves as a reference for the revaluated stresses from the Glinka \textcircled{\smaller B} and Neuber \textcircled{\smaller C} models. 

The simulations show that both approximations \textcircled{\smaller B} and \textcircled{\smaller C} reproduce the stress state very well qualitatively. This is true for both the plane-stress conditions at the surface of the specimen and the more confined stress state in the middle. Both the stress state at the crack tip (\textbf{A}) and in the rounded notch (\textbf{B}) match, even for different load levels for which a varying extend of plasticity can be expected. However, quantitatively, both approximation approaches underestimate the stress.

\begin{figure}
	\def\svgwidth{\linewidth}\small{
		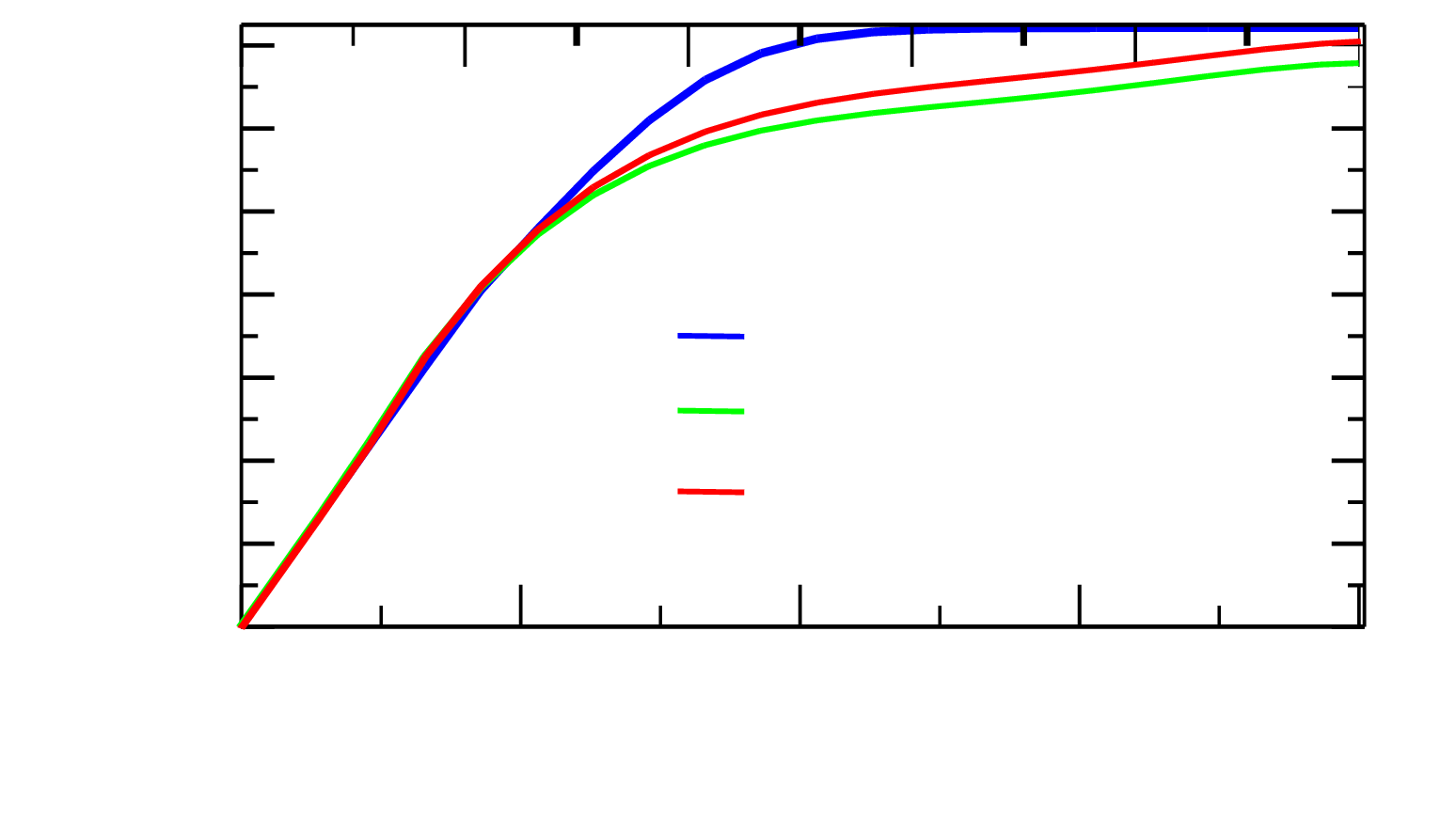}
	\caption{Comparison of equivalent stress for elastic-plastic material model and stress approximations according to Glinka and Neuber. A \underline{monotonic} displacement $\bar{u}$ is applied. MT geometry B without initial crack is used. The fracture toughness is set to a high value to prevent degradation. Stresses are plotted for the point of highest stress in the notch. The approximation models slightly underestimate the stress, depending on the load level.
	}\label{fig:stressplot}
\end{figure}

This can be studied in more detail in Fig.~\ref{fig:stressplot}, which displays the development of the (equivalent) stress over the applied displacement for geometry \textbf{B} at a material point right in the notch. Both stress approximations moderately underestimate the stress determined with the elastic-plastic model. The deviation can be explained with two reasons that Glinka themselves \cite{glinka_energy_1985} already mentioned: For one, the revaluation is carried out for the  Mises equivalent stress, a differentiation between the stress components does not take place.
A greater influence can be expected from the missing redistribution of stresses: During plastification, when the material cannot sustain increasing stresses, they are instead redistributed into the surrounding area -- a process the strictly local stress revaluation based on an elastic solution cannot replicate.

As can be expected from the hypothesis underlying the revaluation, the Glinka model yields slightly lower stresses than the Neuber model. Glinka supposes a more accurate approximation of the SED compared to the elastic results and is therefore the more physically sound model. The difference between the two approximations becomes more pronounced for higher stresses, since the nonlinearity of the stress-strain relation plays a bigger role in that case. The higher deviation of the Glinka model from the elastic plastic model, however, can be traced back to the shortcomings of stress revaluation in general, which is true for both Glinka and Neuber ansatz. Naturally, the quality of the revaluation is closely related to the accuracy of the CSSC, which determines the stresses in the end. As mentioned above, this relation changes over the course of load cycles and the fatigue life of the material.

In fact, the results presented here confirm Glinkas statement that the model underestimates stresses and strains slightly, but that the error in stress is usually not higher than $15\,\%$ -- given that plastic zones are small. 

\subsubsection*{Plastic zone}

\begin{figure}
	\def\svgwidth{\linewidth}\small{
		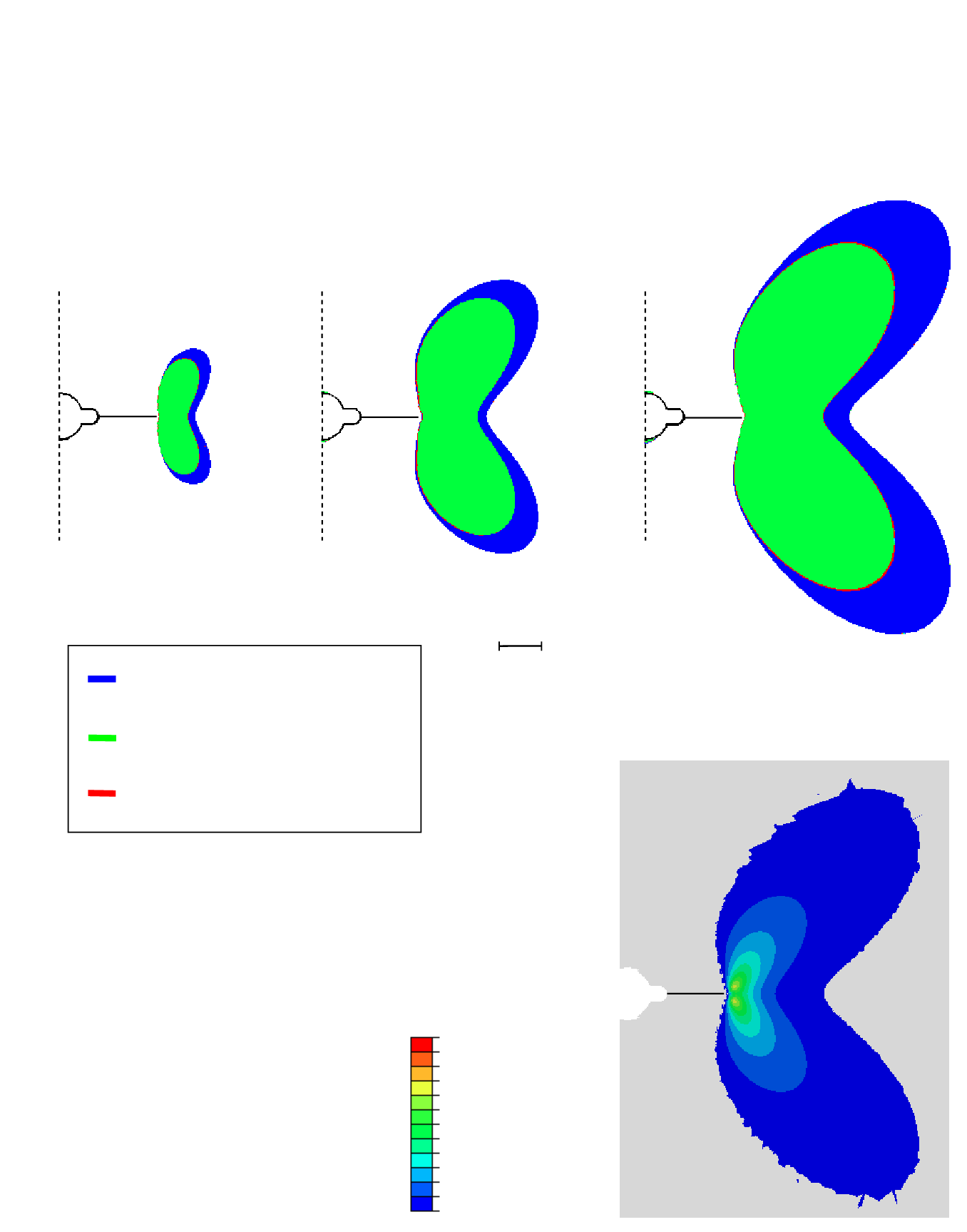}
	\caption{Comparison of plastic zones determined with stress revaluation according to Neuber and Glinka and an elastic-plastic material model. A \underline{monotonic} displacement $\bar{u}$ in tensile direction is applied to an MT specimen. Initial condition B (initial phase-field crack) is adopted. To prevent further cracking, $\Gc$ is set to a high value. Displayed is the plastic zone at the surface as the area in which the (revaluated) equivalent stress surpasses the yield stress $\sigma_0$. Plastic zones according to Neuber (red) and Glinka (green) are almost identical.
		The equivalent plastic strain $\varepsilon^p_\mathrm{eq}$ for the elastic-plastic model at $\bar{u}=0.8$ mm is displayed as well. Please note that this is a numerical experiment not necessarily showing realistic dimensions of the plastic zone.
	}\label{fig:plastZone}
\end{figure}

Fig.~\ref{fig:plastZone} compares the plastic zones for three different load levels for geometry \textbf{A}. Neuber and Glinka produce very similar plastic zones which show the same butterfly shape as the conventional model. The difference in size is in an acceptable range. The plastic zone is identified by the area where the revaluated stress exceeds the initial yield stress $\sigma_0=196.14$\,MPa. Displayed is the plastic zone at the surface. There is no significant difference to the middle of the specimen. For reference, the equivalent plastic strain at $\bar{u}=0.8$\,mm determined by model \textcircled{\smaller A} is shown, too.

Again, the difference between the pseudo-plastic and the elastic-plastic model can be explained by the redistribution of stresses: In the elastic-plastic model, the plastic zone grows larger than  could be expected from an elastic simulation. As the high peak stress of the elastic case is distributed into a larger area of moderate stresses, the area with superseded yield limit is naturally larger. This is an effect the stress revaluation cannot cover, since it is of local nature. That results in a smaller plastic zone compared to the  elastic-plastic model. In fact, the pseudo-plastic zone's size is equal to the hypothetical elastic case here.

In summary, the approximation of the stress state and the plastic zone by the Glinka and also the Neuber model is in an acceptable range. However, the quality of approximation is better for smaller extends of plasticity and strongly depends on the CSSC fed into the model.



\subsection{Plastic contribution to crack driving force}
\label{sec:driving}

Next, we investigate the role of the plastic contribution to the crack driving force and whether its omission in models \textcircled{\smaller B} and \textcircled{\smaller C}  can be justified. 

\subsubsection*{Comparison of crack driving forces}

\begin{figure*}
	\def\svgwidth{\linewidth}\small{
		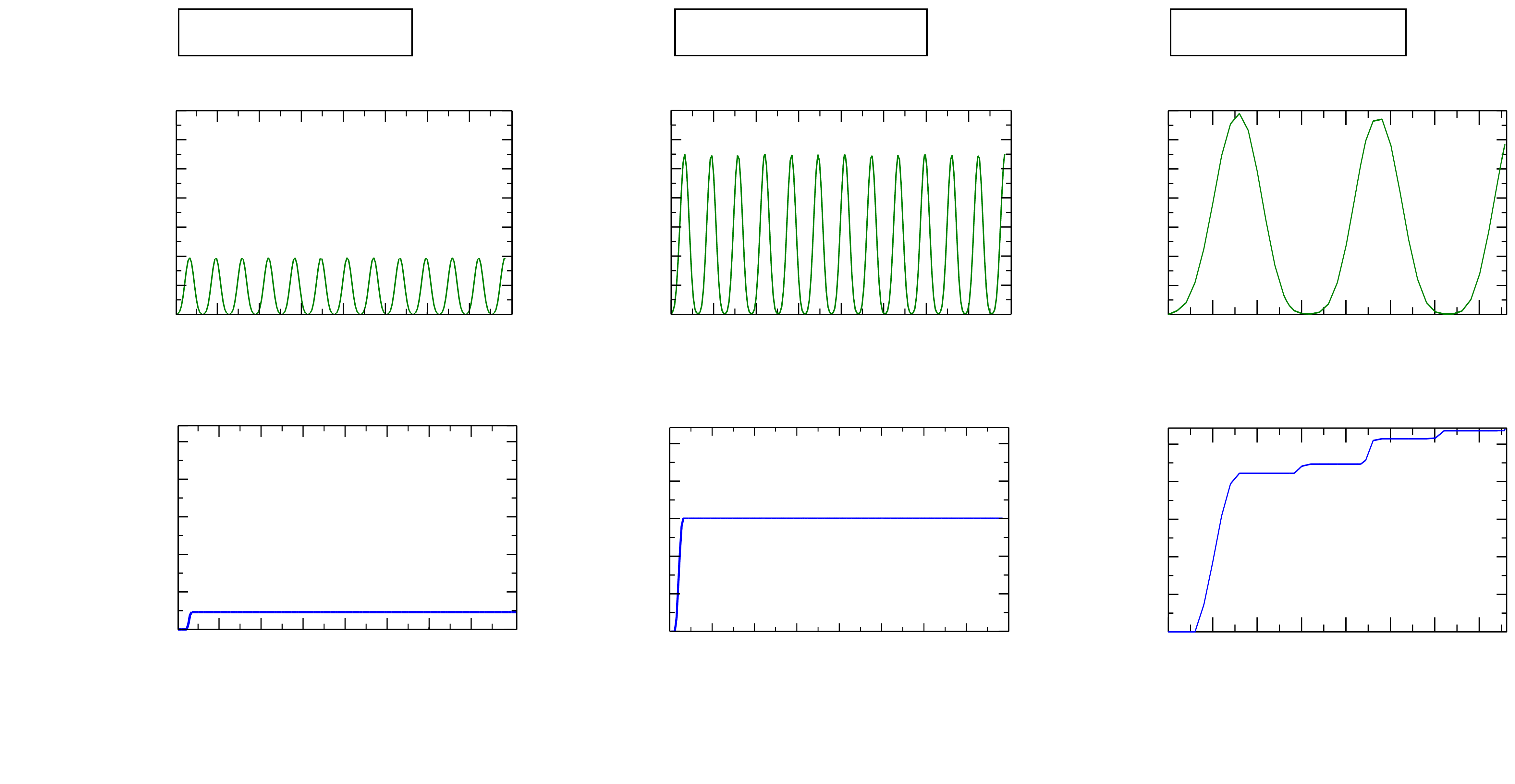}
	\caption{Model \textcircled{\smaller A}: Elastic and plastic contribution to the crack driving force $\psie_+$ and $\psip$. MT geometry \textbf{A} with phase-field initial crack is loaded with {cyclic} displacement load with varying maximum displacement $u_{\max}$. The fracture toughness is set to a high value to prevent further crack propagation. The driving force is displayed for the point of maximum stress within the plastic zone in front of the crack tip. The ratio $\psie_+/\psip$ becomes smaller with higher amplitudes, indicating greater influence of the plastic processes on the fracture behavior. 
	}\label{fig:driving}
\end{figure*}

To this end, we simulate the MT geometry with an initial crack \textbf{A} -- to instantly create a crack tip situation -- with the elastic-plastic model \textcircled{\smaller A}. Again, further crack development is inhibited by a high fracture toughness. An oscillating displacement with varying amplitude is applied. Fig.~\ref{fig:driving} shows the development of elastic and plastic crack contribution to the driving forces $\psie_+$ and $\psip$ at a material point within the plastic zone ahead of the crack tip. 

Naturally, $\psie_+$ oscillates with the applied load. $\psip$, by contrast, increases only when the plastic strain state changes. For constant amplitude loads like in this example, after the first load cycle, this only occurs due to the interplay of kinematic and isotropic hardening. In any case, the ratio between elastic and plastic contribution $\psie_+/\psip$ decreases with increasing load amplitude. While for $u_{\max}=0.2$\,mm it is still 23, it decreases to 6.3 when the load is doubled. This shows that the plastic contribution to the crack driving force can indeed be negligible for smaller loads, but is significant for higher ones.

\subsubsection*{Impact on fatigue life}

\begin{figure}
	\def\svgwidth{\linewidth}\small{
		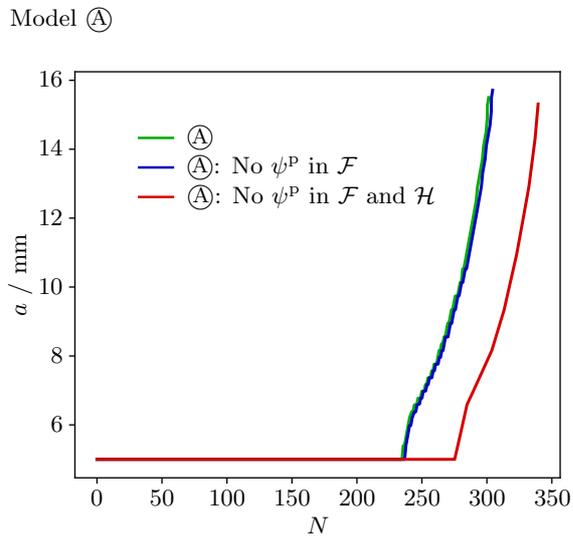}
	\caption{Fatigue crack growth for $R=0.1$ and maximum force $F_{\max}=19.2$\,kN. Displayed is crack length $a$ over load cycle number $N$. Original model \textcircled{\smaller A} is compared to the same with elimination of $\psi^\mathrm{p}$ in fatigue variable $\calF$ and the crack driving force $\calH$. The plastic contribution to the fatigue variable has minor influence on the crack growth rate, in contrast to the plastic crack driving force. 
	}\label{fig:noFpl}
\end{figure}

Which impact the driving forces have on the fatigue life is again studied in with model \textcircled{\smaller A} in Fig.~\ref{fig:noFpl}. To this end, a constant force amplitude with $R=0.1$ and $F_{\max}=19.2$\,kN is applied to geometry \textbf{B}. This time, the fracture toughness is set to its real value, allowing for crack growth. Without the plastic contribution to the crack driving force and the fatigue variable (red curve), crack initiation takes place later than for the unaltered elastic-plastic model (green). For comparison, the model version  with plastic crack driving force, but no plastic contribution to the fatigue variable is shown, too. Its behavior is very similar to the original model \textcircled{\smaller A}, so the difference of the former comparison can be mainly  attributed to the composition of the crack driving force. This underlines again that for certain scenarios, the plastic contribution to the crack driving force can play an important role, so its omission in models \textcircled{\smaller B} and \textcircled{\smaller C} should be undertaken with caution.


\subsubsection*{Remark}

The study in Fig.~\ref{fig:noFpl} is not suitable to draw conclusions on the choice of fatigue variable in the pseudo-plastic models. Generally, its character is very different from the simple accumulated SED employed as a fatigue variable in the elastic-plastic model. Even though the pseudo-plastic model \textcircled{\smaller B} also characterizes the hysteresis by an elastic and a plastic part by the means of a damage parameter,  this parameter is then evaluated with a Wöhler curve and is therefore inherently different from the one in model \textcircled{\smaller A}. Thus, the quality of this modeling choice can only be evaluated long-term by applying the concept to different components and loading scenarios and studying the reproducibility of experimental results for a wider range of applications.

\subsection{Overloads}

Finally, we test the models' ability to deal with more complex loading histories, in this case singular loading events. Experiments show that short sequences of high loads compared to the otherwise dominating load level -- \textit{overloads} -- first lead to a sudden jump in crack growth rate. However, in the following load cycles, the extensive plastification due to the overload causes a retardation in the crack growth rate \cite{bathias_fatigue_2010}. Unusually large plastic strains "shield" the crack in their wake. 

\subsubsection*{Set-up}

In that scenario we compare models \textcircled{\smaller A} and \textcircled{\smaller B}. To this end,  a constant amplitude loading with $F_{\max}=19.2$\,kN and $R=0.1$ is applied to MT geometry \textbf{B}. An overload of 28.8\,kN is applied in between. 

In this case, we aim at a qualitative, not quantitative comparison of the two models' responses to the overload. To be in a feasible load cycle range with model \textcircled{\smaller A}, the scaling factor for the fatigue variable $\vartheta_N$ is again set to 1, causing the crack to propagate at a much larger rate than model \textcircled{\smaller B}. For model \textcircled{\smaller A}, a single overload is applied after 382 load cycles. Model \textcircled{\smaller B} is in a different cycle range, so a block of 7000 overload cycles is applied after 220\,000 lower ones, which gives the same ratio of cycles between both phases.
In order to study significant plastic effects despite the moderate load level, the initial yield stress in \textcircled{\smaller B} is lowered to $\sigma^0=100$\,MPa. $\ell$ is set to 0.5\,mm. All other model parameters are chosen as given in Tab.~\ref{tab:pars}.

\begin{figure*}
	\def\svgwidth{\linewidth}\small{
		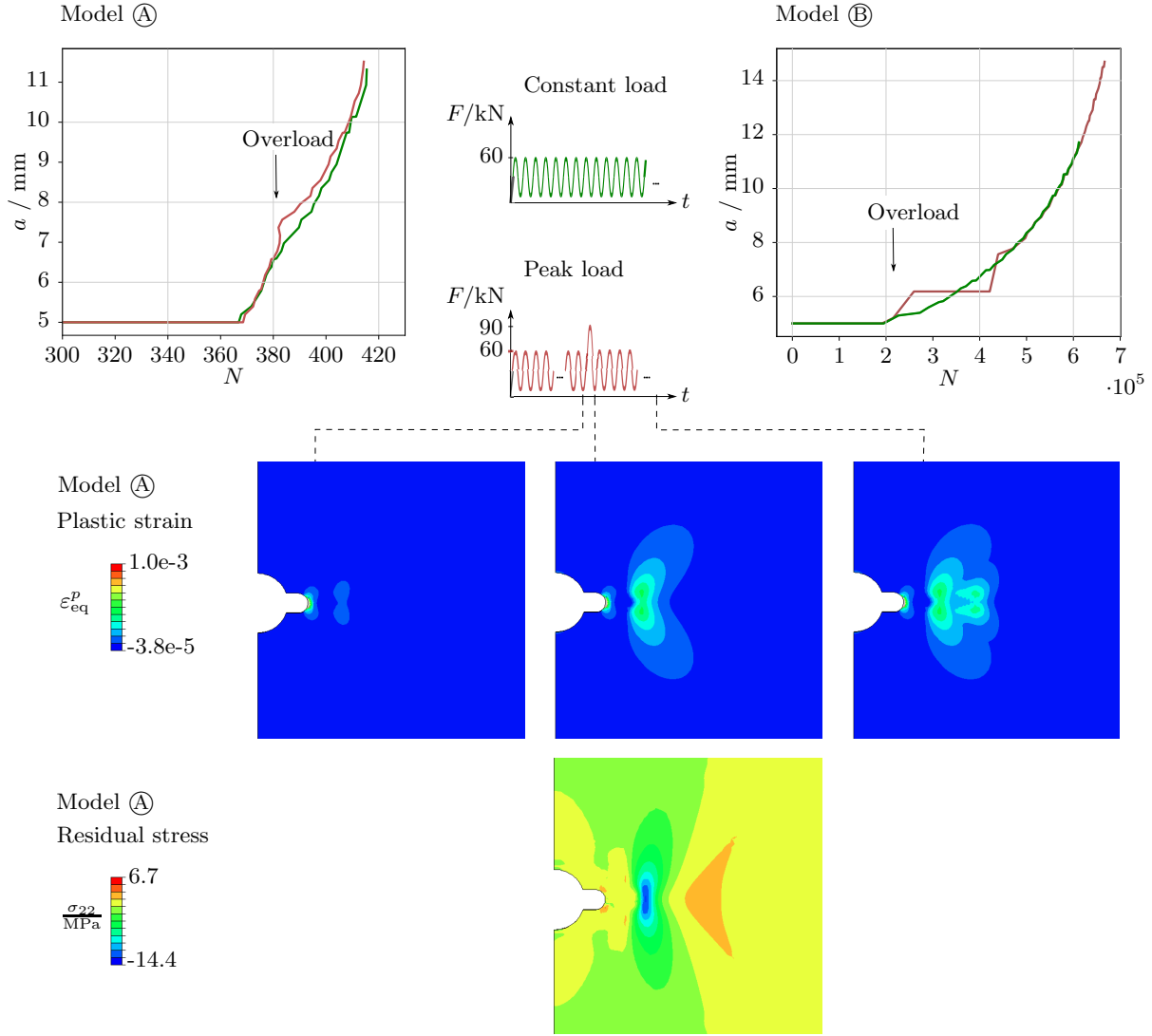}
	\caption{Study of effect of overloads on models \textcircled{\smaller A} and \textcircled{\smaller B}. Constant amplitude loading with $F_{\max}=19.2$\,kN and $R=0.1$ is applied, with or without interruption by an overload of 28.8\,kN. Load cycle numbers differ for both models. After a quick increase in the crack growth rate, both models show a retardation of the crack in the wake of the overload. For model \textcircled{\smaller A}, this is caused by plastic shielding. The plastic zones before and after the overload are depicted below with using the equivalent plastic strain $\varepsilon^\mathrm{p}_\mathrm{eq}$ of von Mises type. Compressive residual stresses caused by the overload also contribute to the retardation, here the crack opening component $\sigma_{22}$ is depicted. For model \textcircled{\smaller B}, all effects can be traced back to the spacial distribution of the fatigue variable. Please note that these numerical experiments do not necessarily capture loading ranges of real experiments.
	}\label{fig:Ueberlast}
\end{figure*}

The results are displayed in Fig.~\ref{fig:Ueberlast}. Both models show an increase in crack growth rate right when the overload is applied compared to the constant-amplitude reference load. Afterwards, as expected, the crack is retarded, albeit in differing manner. 

\subsubsection*{Influence of residual stresses and hardening}

Before evaluating these results, the physical effects causing this retardation effect shall be discussed. According to \cite{bathias_fatigue_2010}, the observed crack growth rate  is reduced until the cyclic plastic zone at the crack tip has left the larger plastic zone caused by the overload. The main reasons for this are strain hardening and residual stresses. Tensile overloads as in this example lead to compressive residual stresses, which reduce the effective stress experienced by the material after the overload.  \cite{waseem_phase_2024} also name plasticity and roughness induced crack closure effects as reasons for the retardation. Typically,  this effect even extends the total fatigue life of the specimen compared to constant amplitude loading. 
Most modeling of the phenomenon is not yet satisfactory, as \cite{bathias_fatigue_2010} state, caused by the great complexity of the process. 

\subsubsection*{Ability of the models to reproduce effect}

While the elastic-plastic model \textcircled{\smaller A} surely does not capture the phenomenon entirely, it incorporates two important aspects the novel model \textcircled{\smaller B} does not: Firstly, it can replicate the formation of residual stresses.  Inhomogeneous distributions of plastic strains are indispensable for that. Fig.~\ref{fig:Ueberlast} exemplarily shows the residual stress component $\sigma_{22}$ right after the overload.
And secondly, it includes the actual hardening phenomena. Isotropic hardening widens the elastic range once the yield limit has been surpassed. In this way, a singular plastic event can prevent plasticity in the next, lower load cycles. The diminished plastic contribution to the fatigue variable in these load cycles adds to the retardation. In Fig.~\ref{fig:Ueberlast}, the equivalent plastic strain marks the plastic zones directly before and after the overload and at the end of the simulation. Clearly, the plastic zone caused by the overload is large enough to influence crack propagation way after the actual overload.

On the contrary, the pseudo-plastic model \textcircled{\smaller B} includes no such effect of sequence. Locally, each load cycle is evaluated anew regarding its contribution to the fatigue variable, independent of former load cycles. Here, the retardation seen in the simulation is very likely caused by the distribution of fatigue damage. The high static overload causes a sudden progress of the crack. Now, at the new crack tip, the fatigue variable has to grow significantly again, before the crack can propagate once more.  However, this delay is caught up soon and has no effect on the overall fatigue life.

Of course there are effects even model \textcircled{\smaller A} does not cover, such as  roughness-caused crack closure. This might also be the reason why the model does not predict an overall extended fatigue life due to overloads. To overcome that, \cite{waseem_phase_2024} e.\,g. deliberately reduce fatigue damage accumulation after the peak load. But even though model \textcircled{\smaller A} does not capture the change in lifetime in this example, it still points out the capacity of an elastic-plastic model to reproduce important plastic effects occurring during complex loading scenarios the pseudo-plastic model \textcircled{\smaller B} is lacking.


\subsection{Conclusions on scope of application}

Based on theoretical considerations on the model structure an the numerical studies, we can now draw conclusions on the scope of application for the novel model \textcircled{\smaller B}. We show that for sufficiently small loading, the model is {capable} to:
\begin{itemize}
	\item Approximate stress distribution at notches and the crack tip sufficiently accurate.
	\item Mimic the shape and size of the plastic zone satisfactorily.
	\item Approximate the crack driving force.
	\item Consider the influence of mean stress through a damage parameter.
	\item Cover crack initiation and propagation. In case of anisotropy, even crack inclination, as shown in \cite{kalina_fatigue_2023}.
	\item Save computational time up to three orders of magnitude compared to elastic-plastic model.
\end{itemize}
The {limitations} of the model lie in modeling of: 
\begin{itemize}
	\item Stress redistribution due to plasticity, especially at the crack tip.
	\item Formation of residual stresses. It requires plastic strains to keep them in place.
	\item Effects of sequence and complex loads. Locally, no history is considered during the computation of fatigue damage.
\end{itemize} 
Considering the aforementioned points, we can summarize the crucial factors to consider when deciding on a model to employ for a specific simulation task:
\begin{itemize}
	\item Size of the plastic zone. This is mainly controlled by the load level and the material's disposition to plastic yielding.
	\item Complexity of the load sequence, i.\,e. overloads and random loading.
	\item Experiments at hand for parametrization. Is cyclic material data available?
	\item Computational time at disposal. The model choice is always a compromise between computational cost and accuracy.
\end{itemize}
In summary, model \textcircled{\smaller B} is ideal for mid-range cyclic loads with small plastic zones. The more load cycles are simulated, the higher the gain in computational time is compared to model \textcircled{\smaller A}: Contrary to most models, the computational effort does not depend greatly on the amount of load cycles. In this way, the speed-up compared to conventional models can be several orders of magnitude at an acceptable level of accuracy.

\section{Summary} 

Balancing between accuracy and computation time, we develop a phase-field model for fatigue fracture with a simplified approach to elastic-plastic material behavior. To this end, we first implement a conventional elastic-plastic phase-field model for fatigue fracture with a plasticity model of Armstrong-Frederick type. The plastic strain energy density contributes to both the crack driving force and the fatigue variable. The model can be expected to yield an accurate description of plastic processes in areas of stress concentrations like the crack tip plastic zone. However, the computational effort is immense, as every load cycle has to be resolved in several increments. This calls for more efficient modeling techniques.

When developing a more efficient model, the  elastic-plastic model serves both as a reference and the point of departure. It is modified in several ways in order to simplify it, yielding a so-called pseudo-plastic model: The stress-strain path is determined approximately with the help of the Glinka revaluation ansatz and additional cyclic material data. This allows for the base of the model to be merely elastic. The plastic strain energy density is omitted in both the crack driving force and during the accumulation of the fatigue variable. In this way, plasticity is considered only in an indirect way through the fatigue variable. Through the omission of a plastic material law, no local iteration is necessary within the material routine. Instead, the elastically computed stress-strain state is revaluated to a pseudo-plastic state. With the stress-strain path known, fatigue damage can be computed easily, again incorporating material data on fatigue resistance. It also allows to skip cycles during the simulation, if crack growth rates are small.

In fact, the novel model is an advancement of a former simplified fatigue model, but with more refined stress-strain revaluation and procedure for the fatigue variable.  We show that the novel model has a significant speed-up of up to two orders of magnitude in simulation time compared to the elastic-plastic model. Moreover, we compare it to both the former model version and the  elastic-plastic model in terms of stress state and crack driving force. In this way, we show that the simplifications lead to tolerable deviations. This is only true for small plastic zones and low load levels, though. Beyond the model's scope are load redistribution due to plasticity and the formation of residual stresses. The decision on the application of the simplified or full model should therefore always depend on the magnitude and complexity of the load and loading path. 

\section*{Acknowledgements}

This work was supported by the Deutsche Forschungsgemeinschaft (DFG) via the project \textit{Experimental analysis and phase-field modelling of the interaction between plastic zone and fatigue crack growth in ductile materials under complex loading} (grant number KA 3309/12-2). The authors are grateful to the Centre for Information Services and High Performance Computing (ZIH) of TU Dresden for providing its facilities for high throughput calculations. 

\section*{Highlights} 

\begin{enumerate}
	\item Simplified phase-field model for ductile fatigue fracture 
	\item Approximation of stress-strain behavior with cyclic experimental data
	\item Comparison to conventional elastic-plastic model
	\item Savings in computation time of up to several order of magnitude compared to conventional model
\end{enumerate}

\bibliography{mybibfile}

\end{document}